\documentclass[%
 reprint,
 superscriptaddress,
 amsmath,amssymb,
 aps,
 pra,
 floatfix,
 longbibliography
]{revtex4-2}

\usepackage{graphicx}
\usepackage{float}
\usepackage{dcolumn}
\usepackage{bm}
\usepackage{lipsum}
\usepackage{braket}
\usepackage{siunitx}
\usepackage{epigraph}
\usepackage{todonotes}
\usepackage[colorlinks]{hyperref}

\begin{document}

\title{Junction-Intrinsic Dissipation in Hybrid Superconductor–Semiconductor Gatemon Qubits}

\author{Zhenhai Sun}
\email{zhenhai.sun@nbi.ku.dk}
\affiliation{NNF Quantum Computing Programme, Niels Bohr Institute, University of Copenhagen, Denmark}

\author{David Feldstein-Bofill}
\affiliation{NNF Quantum Computing Programme, Niels Bohr Institute, University of Copenhagen, Denmark}

\author{Ksenia Shagalov}
\affiliation{NNF Quantum Computing Programme, Niels Bohr Institute, University of Copenhagen, Denmark}

\author{Amalie T. J. Paulsen}
\affiliation{NNF Quantum Computing Programme, Niels Bohr Institute, University of Copenhagen, Denmark}

\author{Casper Wied}
\affiliation{NNF Quantum Computing Programme, Niels Bohr Institute, University of Copenhagen, Denmark}

\author{Shikhar Singh}
\affiliation{NNF Quantum Computing Programme, Niels Bohr Institute, University of Copenhagen, Denmark}

\author{Brian D. Isakov}
\affiliation{Department of Electrical, Computer \& Energy Engineering, University of Colorado Boulder, Boulder, CO 80309, USA}

\author{Jacob Hastrup}
\affiliation{NNF Quantum Computing Programme, Niels Bohr Institute, University of Copenhagen, Denmark}

\author{Christopher W. Warren}
\affiliation{NNF Quantum Computing Programme, Niels Bohr Institute, University of Copenhagen, Denmark}

\author{Svend Krøjer}
\affiliation{NNF Quantum Computing Programme, Niels Bohr Institute, University of Copenhagen, Denmark}

\author{Anders Kringhøj}
\affiliation{NNF Quantum Computing Programme, Niels Bohr Institute, University of Copenhagen, Denmark}

\author{Andr\'as Gyenis}
\affiliation{Department of Electrical, Computer \& Energy Engineering, University of Colorado Boulder, Boulder, CO 80309, USA}
\affiliation{Department of Physics, University of Colorado Boulder, Boulder, CO 80309, USA}

\author{Morten Kjaergaard}
\email{mkjaergaard@nbi.ku.dk}
\affiliation{NNF Quantum Computing Programme, Niels Bohr Institute, University of Copenhagen, Denmark}

\date{\today}

\begin{abstract}
Superconducting transmon qubits based on hybrid superconductor–semiconductor Josephson junctions (gatemons) offer gate tunability, but their relaxation times remain well below those of state-of-the-art transmons, and the origin of this discrepancy is not fully understood. 
Here, we co-fabricate gatemons and SIS-junction transmons with nominally identical circuit layouts, gate dielectrics, and control lines, so that the Josephson element is the only intentional distinction. 
Across multiple chips, transmons in this architecture reach relaxation times in the tens of microseconds, whereas gatemons saturate in the few-microsecond range. 
Using the transmons as on-chip references, we construct a loss budget including Purcell decay, spontaneous emission through the control line, and internal dielectric loss, and find that the corresponding $T_1$ limits exceed all measured gatemon values by more than an order of magnitude. 
Temperature-dependent $T_1$ measurements follow a common quasiparticle-activation model and yield similar superconducting gaps for S–Sm–S and SIS junctions, indicating that the reduced gatemon coherence is dominated by additional temperature-independent, junction-intrinsic dissipation. 
\end{abstract}

\maketitle

\section{Introduction}
Superconducting circuits are a central platform for quantum information processing, quantum sensing, and quantum-limited amplification~\cite{nakamura1999CPB, Schoelkopf2013science, Kjaergaard_2020ARCMP, danilin2021quantumsensingsuperconductingcircuits, Wolski2020PRL, Siddiqi2015Science, wang2025highefficiencylowlossfloquetmodetraveling, Lehnert2017PRX, white2015travelingwaveparametricamplifier}. 
In these circuits, Josephson junctions provide the nonlinearity required to define and control qubits. 
To date, the leading superconducting qubit architectures are based on conventional superconductor–insulator–superconductor (SIS) Josephson junctions. 
Building on this platform, state-of-the-art transmon qubits now achieve relaxation times in the millisecond regime~\cite{bland2025Nat, Bizn_rov__2024NPJ, Sivak_2023Nat, Place_2021NC, berritta_arxiv2025}.
These advances rely on mature junction fabrication, careful control of dielectric participation, and systematic mitigation of known loss channels~\cite{Siddiqi_2021Engineering, Wang_2015AIP}.

An alternative route employs hybrid superconductor–semiconductor–superconductor (S–Sm–S) junctions. 
In such junctions, Andreev bound states with transmission probabilities $\tau_i$ give rise to a nonsinusoidal energy–phase relation
\begin{equation}
E(\phi) = -\Delta \sum_i \sqrt{1 - \tau_i (V_g)\sin^2(\phi/2)},
\end{equation}
where $\Delta$ is the superconducting gap and $V_g$ is a gate voltage that modifies the density of states at the semiconductor part of the junction. 
This enables in-situ gate tunability of the Josephson energy and access to rich mesoscopic physics using circuit QED techniques.
Hybrid junctions have been used to realize gate-tunable transmons (gatemons)~\cite{Larsen_2015, Lange_PRL2015, Casparis_2018}, fluxonium qubits~\cite{Strickland_2025}, and Andreev spin qubits~\cite{Hays_2021, Pita_Vidal_2023}.

Despite this promise, gatemons have so far exhibited relaxation times $T_1$ that are systematically shorter than those of state-of-the-art transmons.
The best reported gatemons reach $T_1 \sim \SI{30}{\micro\second}$~\cite{Luthi_PRL2018, purkayastha_arxiv2025}, while most devices show $T_1 \lesssim \SI{10}{\micro\second}$~\cite{Casparis_PRL2016, Kringhoj_PRApplied2021}; in stark contrast to modern transmon based devices.
The physical origin of this performance gap remains unclear. Previous gatemon studies have primarily focused on device demonstration, gate tunability and stability, and integration into cQED architectures~\cite{Larsen_2015, Lange_PRL2015, Casparis_2018, Hertel_2022PRApplied, Sagi_2024NatComm, Riechert_2025NatComm, Zheng_NanoLett2024, David_25PRApplied}, while systematic investigations of relaxation mechanisms are scarce~\cite{Strickland_PRR2024}. 
In contrast, extensive work on SIS-based transmons has identified dominant loss channels such as Purcell decay~\cite{Houck_08PRL}, dielectric loss associated with substrates and interfaces~\cite{Siddiqi_2021Engineering, Wang_2015AIP}, and quasiparticle poisoning~\cite{Catelani_2011PRB, Catelani_12PRB, Kyle_18PRL}.

A direct, controlled comparison between gatemons and transmons that share the same materials stack, circuit layout, and control-line architecture is therefore needed to disentangle junction-specific dissipation from losses associated with substrate, interfaces, or fabrication flow. 
In particular, hybrid S–Sm–S junctions may exhibit a smaller induced gap than the parent superconductor~\cite{Chang_NatNano2015, Valentini_NatComm2024} and finite subgap conductance~\cite{Liu_PRB2017, Sarma_PRL2013, Levajac_NatComm2023}, both of which can modify quasiparticle dynamics and could introduce additional relaxation channels. 
Isolating these effects from more standard superconducting qubit loss mechanisms is essential for assessing the ultimate coherence potential of hybrid superconductor–semiconductor qubits.

In this work, we investigate the persistent discrepancy that gatemon qubit relaxation times remain significantly shorter than those of state-of-the-art transmon qubits.
To isolate the cause of this difference, we co-fabricate gatemon and transmon devices on the same high-resistivity silicon chips with identical circuit designs, fabrication steps, and gate dielectric layers (Fig.~\ref{fig:Figure1_optical}(a–c)), with the only intentional distinction being the Josephson junction.
We implement a gatemon-optimized XYZ-style control-line architecture~\cite{Manenti_APL2021} for both qubit types and design the gatemon gate layout to reconcile strong DC tunability with suppressed spontaneous emission. 
Within this co-integrated platform, we systematically compare $T_1$ across multiple chips and material stacks, construct a loss budget using the transmons as on-chip references, and perform temperature-dependent $T_1$ measurements on paired devices. 
We find that Purcell decay, gate-line spontaneous emission, internal dielectric loss, and thermally activated quasiparticles cannot account for the short gatemon $T_1$. 
Instead, our results point to additional, temperature-independent dissipation intrinsic to the S–Sm–S junction.

\section{Co-design of gatemon and transmon qubits} \label{sec:gatemontransmon}

\begin{figure}[t!]
    \centering
    \includegraphics[width=\columnwidth]{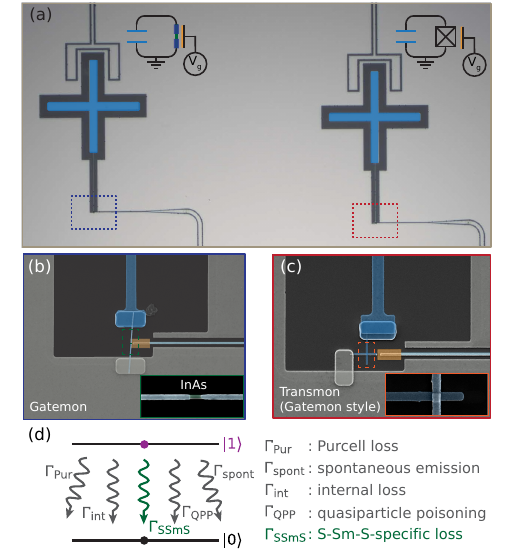}
    \caption[Images of the device]{(a) Optical image of the device containing both gatemon (left) and transmon (right) qubits. 
    Meandered readout resonators are capacitively coupled to the qubit islands for dispersive readout. 
    The transmons and gatemons share an identical circuit layout, including the X-shaped qubit capacitor and drive/gate lines and fabrication steps, ensuring a comparable electromagnetic environment. 
    The zoom-in images of the Josephson junction areas, highlighted in blue and red, are shown in (b) and (c) for the gatemon and transmon, respectively. 
    (b) The gatemon employs a superconductor–semiconductor–superconductor (S–Sm–S) hybrid Josephson junction based on InAs/Al nanowires. One segment of the Al shell is selectively etched to form the S–Sm–S weak link, as shown in the inset. 
    (c) The transmon uses a conventional superconductor–insulator–superconductor (SIS) Josephson junction. 
    Gate dielectric $\mathrm{HfO_x}$ (orange) is applied to both qubits to compare device performance under as similiar conditions as possible.
    (d) The primary loss mechanisms under investigation: the Purcell effect, spontaneous emission, internal loss, quasiparticle poisoning, and possible S–Sm–S-junction-specific loss, with corresponding relaxation rates labeled as $\Gamma_\mathrm{Pur}$, $\Gamma_\mathrm{spont}$, $\Gamma_\mathrm{int}$, $\Gamma_\mathrm{QPP}$, and $\Gamma_\mathrm{SSmS}$.
    }
    \label{fig:Figure1_optical}
\end{figure}

Our experimental platform employs full-shell InAs/Al nanowire-based gatemons~\cite{Larsen_2015, Anders_PRB2018, Casparis_2018, Casparis_PRL2016} since they can be fabricated on the same high-resistivity silicon substrates with identical process flow to conventional planar transmons. 
This choice embeds both qubit types in the same Si-based, highly optimized transmon technology stack, so that any coherence differences can be attributed primarily to the semiconducting Josephson junction rather than to the surrounding materials or fabrication flow.
By contrast, in other S–Sm–S implementations, such as 2DEG-based junctions, the Josephson element resides in a III–V heterostructure with a different substrate and dielectric~\cite{Strickland_PRR2024, Hertel_2022PRApplied, Sagi_2024NatComm}. 
In those systems, any difference in $T_1$ would inevitably convolve junction-specific loss with changes in substrate and interface loss, obscuring the junction’s intrinsic contribution. 
In our nanowire implementation, co-fabricated transmons on the same chip serve as on-chip reference qubits that share the same materials stack, circuit geometry, and control lines, suppressing chip-to-chip variability and providing a benchmark that isolates loss channels specific to the S–Sm–S gatemon junction.

Figure~\ref{fig:Figure1_optical}(a) shows an optical image of a representative sample from our study.
Fabrication details are provided in Appendix.~\ref{app:C}. 
The exactly identical device layouts of the gatemon and the transmon-with-a-gatemon-design are shown in Fig.~\ref{fig:Figure1_optical}(b) and (c). 
We emphasize that the microwave drive line for the transmon (Fig.~\ref{fig:Figure1_optical}(c)) shares the same geometry and dielectric atomic layer deposition (ALD) stack as the combined drive- and gate-line of the gatemon. 
For the transmon, this line is used purely for microwave control, as the qubit frequency is fixed by a single SIS junction without a SQUID loop, whereas for the gatemon the same layout provides both DC gating and microwave drive. 
Using an identical line geometry and dielectric environment for both qubit types ensures that any observed difference in $T_1$ cannot be ascribed to differences in control-line coupling or spontaneous-emission engineering, but instead reflects the distinct Josephson elements and their coupling to the same environment. 
The detailed optimization of this gate-line geometry, including its XYZ-style~\cite{Manenti_APL2021} implementation for gatemons and the resulting spontaneous-emission limit, is discussed in Sec.~\ref{sec:loss_mechanism} and illustrated in Fig.~\ref{fig:design}. 

\section{Benchmarking of co-integrated gatemons and transmons}
\begin{figure}
    \centering
    \includegraphics[width=\columnwidth]
    {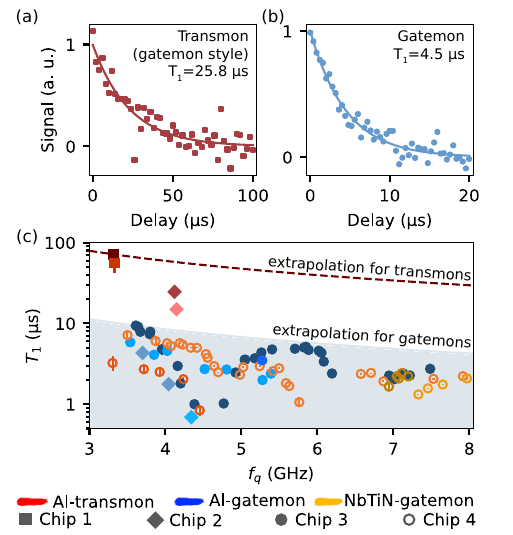}
    \caption{\textbf{Comparison of relaxation times between transmons and gatemons.}
    (a) Energy relaxation measurement of a representative transmon.
    (b) Energy relaxation measurement of a representative gatemon at similar qubit frequencies to the fixed-frequency transmon in panel (a). Both qubits are located on the same chip (Chip 2).
    (c) Summary of $T_1$ times for all measured devices across four different chips. 
    Chips~1--3 are fabricated using aluminum films, and Chip~4 uses a NbTiN film. 
    Qubits on the same chip are indicated by the same marker shape, while different colors are used to distinguish individual qubits. 
    The qubits are grouped into three categories: aluminum transmons, aluminum gatemons, and NbTiN gatemons, each represented by a distinct color palette. 
    The red dashed line shows the extrapolated $T_1$ limit for a transmon with a constant quality factor $Q = 2\pi f T_1$, while the shaded region indicates the expected $T_1$ range for the best-performing gatemon extrapolated across frequencies.
    }
    \label{fig:T1_summary}
\end{figure}

We now turn to the measured energy relaxation times. 
We first characterize the relaxation times $T_1$ of both fixed-frequency transmons and gatemons as a function of qubit frequency $f_q$, with all measurements performed in a dilution refrigerator at a base temperature of approximately $\SI{30}{\milli\kelvin}$.
Representative energy-relaxation traces are shown in Figs.~\ref{fig:T1_summary}(a,b). 
For the transmon, we obtain $T_1=\SI{25.8}{\micro\second}$ at $f_q = \SI{4.21}{\giga\hertz}$, while the gatemon measured on the same platform yields $T_1 = \SI{4.5}{\micro\second}$ at $f_q = \SI{3.70}{\giga\hertz}$.

To move beyond single-device examples, we fabricate and characterize several chips and summarize the resulting $T_1$ values in Fig.~\ref{fig:T1_summary}(c). 
Each data point represents a time average over multiple $T_1$ measurements (at least 15 minutes) to account for temporal fluctuations. 
In total, we study four transmons and five gatemons on three Al-film chips, with two transmons and two gatemons co-located on the same chip (Chip 2). 
We also include three additional gatemons fabricated on a NbTiN film (Chip 4), a material platform that has previously been used for high-coherence transmon~\cite{Bruno_APL2015, Murray_2021, Ani_arxiv2019}.
In the resulting data set, aluminum transmons in this architecture routinely reach $T_1$ values in the tens of microseconds, with the best device achieving $T_1 = \SI{71.6}{\micro\second}$, whereas gatemon $T_1$ values cluster in the few-microsecond range, with a maximum of $T_1 = \SI{9.1}{\micro\second}$ (independent of Al or NbTiN ground plane material), corresponding to roughly a factor-of-six between the best performers in each family.

Within this co-integrated platform, the only intentional difference between gatemons and transmons is the Josephson junction (S–Sm–S nanowire vs. SIS tunnel), while the capacitor geometry, substrate, gate dielectric, control lines, and readout resonators are shared. 
The systematic $T_1$ offset observed in Fig.~\ref{fig:T1_summary}(c) therefore points to additional dissipation channels associated with the hybrid S–Sm–S junction, rather than differences in substrate loss, circuit layout, or control-line engineering.
In the following, we therefore quantitatively evaluate the expected contributions from Purcell decay, spontaneous emission through the control lines, internal dielectric loss, and thermal quasiparticle poisoning, and show that their combined effect cannot account for the observed discrepancy in $T_1$, implying an additional loss channel intrinsic to the S–Sm–S junction.

\begin{figure}[t!]
    \centering
    \includegraphics[width=\columnwidth]{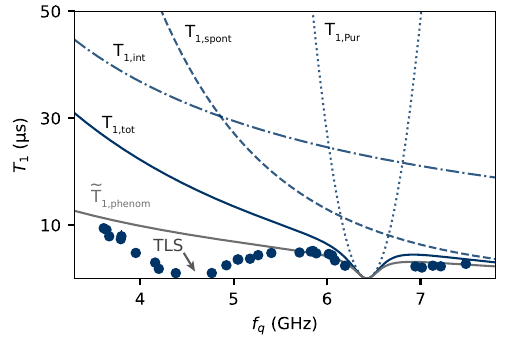}
    \caption[Qubit relaxation time as a function of qubit frequency]{\textbf{Gatemon relaxation time as a function of qubit frequency $f_q$.}
   Dashed lines indicate the expected limits from Purcell decay, spontaneous emission through the gate line, and internal loss. 
    The solid line shows the combined theoretical limit. 
    The pronounced dip near $\SI{6.43}{\giga\hertz}$ corresponds to the readout resonator frequency. 
    The reduction in $T_{1}$ around $\SI{4.5}{\giga\hertz}$ is attributed to a spurious two-level system (TLS), and is not a generic feature of our devices.
    }
    \label{fig:loss_mechanism}
\end{figure}

\section{Loss budget in gatemon qubits} \label{sec:loss_mechanism}
We next analyze the individual contributions to gatemon relaxation by constructing a loss budget for a representative device, as summarized in Fig.~\ref{fig:loss_mechanism}.
We plot the recorded $T_1$ values of a representative device, as a function of qubit frequency $f_q$ (blue circles).
The qubit relaxation time can be expressed in terms of the quality factor
\begin{equation}
Q = 2\pi f_q T_1 = 2\pi f_q \left(\frac{1}{\Gamma_1}\right),
\end{equation}
where $\Gamma_1$ is the total energy decay rate. Assuming independent loss channels, the total quality factor can be written as
\begin{equation}
\frac{1}{Q_\text{tot}}=\frac{1}{Q_\mathrm{Pur}}+\frac{1}{Q_\mathrm{spont}}+\frac{1}{Q_\mathrm{int}}+\frac{1}{Q_\mathrm{QPP}}+ \cdots ,
\label{eq:Qsum}
\end{equation}
where $Q_\mathrm{Pur}$, $Q_\mathrm{spont}$, $Q_\mathrm{int}$, and $Q_\mathrm{QPP}$ denote the contributions from Purcell decay, spontaneous emission through external lines, internal (dielectric and conductor) loss, and quasiparticle poisoning, respectively. In the following, we estimate each term and compare the corresponding limits to the measured gatemon $T_1$.

The Purcell effect arises from enhanced spontaneous emission due to the coupling between the qubit and the readout resonator. In the dispersive regime, the Purcell-limited relaxation time is given by
\begin{equation}
T_{1,\mathrm{Pur}} =
\frac{(\omega_q - \omega_r)^2}{\kappa g_{qr}^2},
\label{eq:Purcell}
\end{equation}
where $\omega_q = 2\pi f_q$ is the qubit frequency, $\omega_r = 2\pi f_r$ is the readout resonator frequency, $\kappa = \omega_r/Q_r$ is the resonator energy decay rate, and $g_{qr}$ is the qubit–resonator coupling strength~\cite{Houck_08PRL}. From our measurements we estimate $\kappa/2\pi \approx \SI{0.87}{\mega\hertz}$ and $g_{qr}/2\pi \approx \SI{35}{\mega\hertz}$. 
Substituting these values into Eq.~\eqref{eq:Purcell} shows that $T_{1,\mathrm{Pur}}$ exceeds $\SI{300}{\micro\second}$ over the range $f_q \in [3,5]~\mathrm{GHz}$, plotted as a dotted lines in Fig.~\ref{fig:loss_mechanism}.

Another dissipation channel arises from capacitive coupling between the gate (or drive) line and the qubit, leading to spontaneous emission into the external control circuitry. 
The associated relaxation time can be written as
\begin{equation}
T_{1,\mathrm{spont}} = \frac{C_\textrm{q}}{\text{Re}[Y(\omega_q)]},
\end{equation}
where $C_\textrm{q}$ is the total qubit capacitance and $Y$ is the admittance to the control line as seen by the qubit~\cite{Neely_PhysRevB2008}.
\begin{figure}[!t]
    \centering
    \includegraphics[width=\columnwidth]{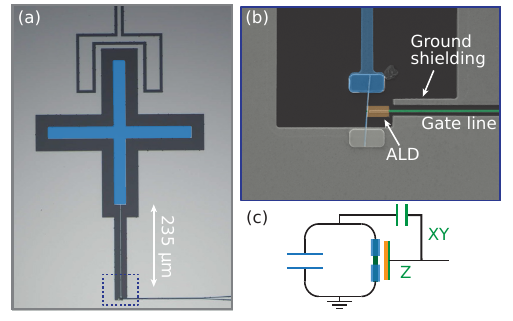}
    \caption{\textbf{Detailed gatemon qubit design.}
    (a) Representative optical image of a gatemon qubit from this work.
    The gate line is placed $\SI{235}{\micro\meter}$ away from the qubit capacitor (blue). 
    Zoom in of the dashed box region is shown in (b). 
    This segment of the gate line is designed to be $\SI{200}{\nano\meter}$ wide. 
    The ground plane is brought along with the gate line, the ground shielding, to shield the qubit from this line. 
    (c) Circuit diagram of the qubit. 
    The single gate line is in close proximity to the junction for effectively tuning the Josephson energy ($Z$ control), and serve as a drive line ($XY$ control) via optimized coupling capacitance to the qubit (see text for details).
    }
    \label{fig:design}
\end{figure}
In gatemons, a gate electrode must already be brought close to the junction to control the carrier density, so it is natural to use the same line for both DC tuning and microwave drive, provided this can be done without introducing a severe $T_1$ penalty. 
Implementing such XYZ-style control is more challenging than in SIS-based transmons, because the gate electrode must be brought into close proximity ($\lesssim \SI{100}{\nano\meter}$) to the junction to efficiently modulate the carrier density in the semiconductor. 
To reconcile strong DC tunability with controlled AC coupling at qubit frequencies, without introducing a prohibitive spontaneous-emission limit on $T_1$, we redesign the gate layout as shown in Figs.~\ref{fig:design}(a-c). 
We implement three modifications to reduce the parasitic capacitance while maintaining gate proximity to the junction. 
First, the distance between the X-shaped qubit capacitor and the gate line is increased to more than $\SI{200}{\micro\meter}$. 
Second, the gate line is patterned to be $\sim \SI{200}{\nano\meter}$ wide near the qubit capacitor. 
Third, the ground plane is brought along with the gate line to shield the qubit from the gate-line electric field, as shown in Fig.~\ref{fig:design}(b). 

With these optimizations, we perform electromagnetic simulations and find a coupling capacitance of $C_c = \SI{0.19}{\femto\farad}$ and frequency-dependent admittance $Y(\omega_\text{q})$. The corresponding spontaneous emission limit is above $\SI{30}{\micro\second}$ over the frequency range $f_\mathrm{q} \in [3, 5]~\mathrm{GHz}$. 
This maintains the relaxation times of the device while preserving the DC control of the carrier density in the nanowire.

The third channel we consider is internal loss originating from dielectric and conductor dissipation in the capacitor, substrate, interfaces, and gate dielectric. 
Because the gatemon and transmon qubits share an identical circuit geometry and materials stack—including the ALD gate dielectric—their electric-field distributions and participation ratios in the various materials are effectively the same. 
This design parity allows us to use the co-fabricated transmons as direct probes of the internal quality factor for this stack. 
At the measured transmon frequencies (around $\SI{3.4}{\giga\hertz}$ and $\SI{4.2}{\giga\hertz}$), the relaxation times are predominantly limited by internal loss, from which we extract an average internal quality factor $Q_\text{int} \approx 0.93\times 10^6$. The corresponding internal-loss limit for the gatemon is then
\begin{equation}
T_{1,\mathrm{int}} = \frac{Q_\text{int}}{2\pi f_q},
\end{equation}
which yields expected lifetimes in the tens-of-microseconds range over the frequencies of interest, shown as dashed-dotted lines in Fig.~\ref{fig:loss_mechanism}.

Combining these three contributions according to Eq.~\eqref{eq:Qsum} gives the total expected relaxation time $T_{1,\mathrm{tot}}$ for the gatemon, shown as the solid line in Fig.~\ref{fig:loss_mechanism}. Across the full frequency range, including away from the narrow dip associated with a strongly coupled two-level system, this predicted lifetime remains well above all experimentally observed gatemon $T_1$ values, demonstrating that Purcell decay, gate-line spontaneous emission, and internal loss are insufficient to account for the measured relaxation and that an additional, yet unidentified loss channel must be present in the S–Sm–S junction.

To quantify how large such an additional loss channel must be, we extend Eq.~\eqref{eq:Qsum} by adding a phenomenological, frequency-independent term $1/Q_{\mathrm{phenom}}$ to the total quality factor. 
The corresponding relaxation time $\widetilde T_{1,\mathrm{phenom}}$ is plotted as the gray curve in Fig.~\ref{fig:loss_mechanism}. 
For the representative device shown, choosing $Q_{\mathrm{phenom}} \approx 2.5\times 10^{5}$ brings the model into good agreement with the measured $T_1$ over the full frequency range, apart from the narrow dip attributed to a strongly coupled spurious two-level system. 
We emphasize that the precise value of $Q_{\mathrm{phenom}}$ is device dependent and not universal; its significance is that a modest additional loss channel, distinct from Purcell decay, gate-line spontaneous emission, and internal dielectric loss, is sufficient to reconcile the data with an otherwise standard loss budget. 
At present we do not assign a specific microscopic mechanism to $Q_{\mathrm{phenom}}$, and instead regard it as an effective parameter capturing junction-intrinsic dissipation that is further constrained by the temperature-dependent measurements in Sec.~\ref{sec:temperature_dependence}. 

\begin{figure}
    \centering
    \includegraphics[width=\columnwidth]{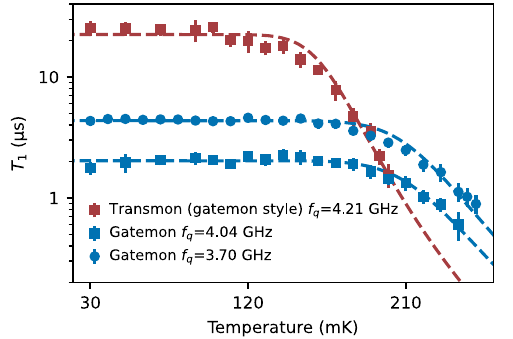}
    \caption{\textbf{Qubit relaxation times $T_1$ as a function of temperature.}
    $T_1$ measurements for a gatemon (blue) and a transmon (red) from Chip 2 at different temperatures.
    The gatemon is measured at two different qubit frequencies, $\SI{3.70}{\giga\hertz}$ and $\SI{4.04}{\giga\hertz}$. 
    At base temperature ($\SI{30}{\milli\kelvin}$), the transmon exhibits a relaxation time of approximately $\SI{25}{\micro\second}$, while the gatemon shows $T_1\approx\SI{4.5}{\micro\second}$ and $T_1\approx\SI{2.7}{\micro\second}$ for the high- and low-frequency configurations, respectively. 
    The dashed lines are fits using a quasiparticle poisoning model, $T_1(T) = \bigl[\Gamma_0 + (\sqrt{8\Delta h f_{\mathrm{q}}}/h)\sqrt{2\pi k_{\mathrm{B}}T/\Delta}\, e^{-\Delta/(k_{\mathrm{B}}T)}\bigr]^{-1}$, where $\Gamma_0$ is the temperature-independent decay rate, $k_\mathrm{B}$ is the Boltzmann constant, and $\Delta$ is the superconducting gap.
        }
    \label{fig:T1_vs_temp}
\end{figure}

\section{Temperature dependence of $T_1$ for co-designed gatemon and transmon}\label{sec:temperature_dependence}
Having established that Purcell decay, gate-line spontaneous emission, and internal loss cannot account for the short gatemon relaxation times, we next investigate relaxation due to quasiparticle poisoning (QPP). 
The presence of nonequilibrium quasiparticles at millikelvin temperatures is a well-known limitation for state-of-the-art transmons, and their influence on qubit relaxation has been extensively studied using, e.g., direct parity-switching measurements~\cite{Sun_PRL2012, Rist_NatComm2013, Kyle_18PRL, Krause_PRApplied2024, Connoly_PRL2024}, intentional quasiparticle injection~\cite{Diamond_PRXQuantum2022, Liu_PRL2024, Benevides_PRL2024, Larson_PRXQuantum2025}, and temperature-dependent $T_1$ measurements~\cite{Kyle_18PRL, Pan_NatComm2022, Martinis_PRL2009, Rist_NatComm2013}. 
Direct qubit parity-switching techniques have recently been applied to gatemons as well~\cite{uilhoorn_arxiv2021, Erlandsson_PRB2023}. Yet, to our knowledge, systematic measurements of $T_1$ as a function of temperature in gatemons have not yet been reported.
Compared with SIS Josephson junctions, S–Sm–S junctions rely on induced superconductivity in the semiconductor~\cite{Beenakker_PRB1992, Aguado_APL2020}, which could, in principle, exhibit a smaller superconducting gap than the parent superconductor. 
Moreover, such hybrid junctions are known to host finite in-gap (subgap) conductance~\cite{Sarma_PRL2013, Chang_NatNano2015, Vaitiek_Science2020}, even when the superconductor–semiconductor nanowires are epitaxially grown for improved interface quality as the nanowire in the present work~\cite{Chang_NatNano2015}. 
Both effects can modify quasiparticle dynamics and introduce additional dissipation channels, motivating a dedicated temperature study.

We perform temperature-dependent measurements on a gatemon and a transmon co-fabricated on the same chip (Chip 2), as shown in Fig.~\ref{fig:T1_vs_temp}. 
The transmon has a fixed frequency of $f_q = \SI{4.21}{\giga\hertz}$, while the gatemon is tuned to two nearby operating points at $f_q = \SI{4.04}{\giga\hertz}$ and $f_q = \SI{3.70}{\giga\hertz}$.
At each temperature, we raise the mixing-chamber setpoint, allow the system to equilibrate, and then record multiple $T_1$ traces for both devices. 
Both qubits exhibit a temperature-independent $T_1$ plateau at low temperatures, followed by a pronounced decrease at elevated temperatures, consistent with thermally activated quasiparticles. The QPP-induced relaxation rate is proportional to the quasiparticle density
\begin{equation}
x_\mathrm{qp}(T)=x_\mathrm{ne}+\sqrt{\frac{2\pi k_\mathrm{B} T}{\Delta}}\, e^{-\frac{\Delta}{k_\mathrm{B}T}},
\end{equation}
where $x_\mathrm{ne}$ is the nonequilibrium quasiparticle population at the base temperature, $k_\mathrm{B}$ is the Boltzmann constant, and $\Delta$ is the superconducting gap. In the limit of $E_\mathrm{J} \gg E_\mathrm{C}$, the temperature dependence of the relaxation time can be modeled as
\begin{equation}
T_1(T) = \frac{1}{\Gamma_\mathrm{0} + \frac{\sqrt{8\Delta hf_\mathrm{q}}}{h}\sqrt{\frac{2\pi k_\mathrm{B} T}{\Delta}}\, e^{-\frac{\Delta}{k_\mathrm{B}T}}},
\label{eq:T1_QP}
\end{equation}
where $\Gamma_0$ collects all temperature-independent decay channels~\cite{Glazman_SciPost2021}. 
Explicitly, $\Gamma_0$ includes residual Purcell decay, internal loss, spontaneous emission through the control lines, relaxation due to nonequilibrium quasiparticles, $\Gamma_\mathrm{0, ne} = x_\mathrm{ne}\sqrt{8\Delta hf_q}/h$, and any additional temperature-independent loss specific to the junction.

Fitting the transmon and gatemon data to Eq.~\eqref{eq:T1_QP} yields $1/\Gamma_0 = \SI{2.03\pm 0.03}{\micro\second}$ and $\Delta/h = \SI{55.2\pm0.3}{\giga\hertz}$ for the gatemon at $f_q = \SI{4.04}{\giga\hertz}$, and $1/\Gamma_0 = \SI{23 \pm 1}{\micro\second}$ and $\Delta/h = \SI{48.8 \pm 0.3}{\giga\hertz}$ for the transmon. 
Thus, the extracted superconducting gap of the S–Sm–S junction is comparable to that of the SIS junction within the uncertainties of the fit, consistent with expectations. 
The temperature dependence of $T_1$ for both devices is well captured by the same quasiparticle-activation form,  indicating that the thermally generated quasiparticles affect the two qubits in a quantitatively similar way. 
These observations suggest that a substantially smaller induced gap in the hybrid junction, or a qualitatively different thermal quasiparticle population, is not the primary origin of the shorter gatemon $T_1$.

The key difference between the two devices is instead encoded in the temperature-independent term $\Gamma_0$. 
At base temperature, the transmon exhibits $T_1 \approx \SI{23}{\micro\second}$, whereas the gatemon saturates at $T_1 \approx \SI{4.6}{\micro\second}$ and $T_1 \approx \SI{2.0}{\micro\second}$ for the two operating points, captured by a significantly larger $\Gamma_0$ for the gatemon.
One possible explanation for the larger $\Gamma_0$ is that the S–Sm–S junction has a substantially higher excess nonequilibrium quasiparticle density.
In the case where $\Gamma_0$ is dominated by nonequilibrium quasiparticles, we can calculate the corresponding upper bound of $x_\mathrm{ne} \sim 10^{-5}$, which is several orders of magnitude higher than the standard transmons.
To directly measure the density of excess quasiparticles, a Ramsey-style experiment with offset-charge-sensitive gatemons can provide more information about the exact number of $x_\mathrm{ne}$ in future work~\cite{Kyle_18PRL}.
Another possible candidate for high $\Gamma_0$ is quasiparticle tunneling mediated by the residual subgap density of states~\cite{Kringhoj_PRApplied2021, Levajac_NatComm2023}, which can originate from the imperfect transparency of the S-Sm interface~\cite{Chang_NatNano2015}.
Those residual low-energy subgap states potentially enhance the quasiparticle tunneling rate, thereby providing another channel for energy dissipation.
Further investigations of subgap states and interface engineering in S–Sm–S junctions will further disentangle these contributions.

\section{Conclusion}
We have carried out a systematic investigation of energy relaxation in hybrid superconductor–semiconductor gatemons using co-fabricated SIS-junction transmons as on-chip references. 
By realizing both qubit types on the same high-resistivity silicon substrates, with nominally identical circuit layouts, gate dielectrics, control lines, and readout resonators, we isolate the Josephson element as the only intentional distinction. 
Across multiple chips and material platforms, transmons in this architecture reach relaxation times in the tens of microseconds, whereas gatemons saturate in the few-microsecond range, revealing a robust gap in $T_1$ between the two qubit families. 

Using the transmons to benchmark standard loss channels, we construct a loss budget for one of our representative gatemons that includes Purcell decay, spontaneous emission through the optimized control line, and internal dielectric loss.
The corresponding limits remain well above all measured gatemon $T_1$ values, showing that these mechanisms cannot explain the observed relaxation.
By augmenting this budget with a single phenomenological, frequency-independent loss term associated with the S–Sm–S junction, we obtain a curve that reproduces a representative gatemon $T_1$, showing that the presence of a modest additional junction-specific loss channel is sufficient to explain the data even though its microscopic origin remains unresolved. 
Temperature-dependent $T_1$ measurements on paired gatemon–transmon devices are described by the same quasiparticle-activation model and yield comparable superconducting gaps for the S–Sm–S and SIS junctions, ruling out a substantially reduced induced gap as the primary source of the shorter gatemon lifetimes. 
The remaining discrepancy appears as a larger temperature-independent decay rate for the gatemon, pointing to additional dissipation channels intrinsic to the hybrid junction. 

A potential interpretation is that relaxation limitations in gatemons arise from junction-intrinsic mechanisms such as finite subgap conductance and imperfect superconductor–semiconductor interfaces, which provide a residual in-gap density of states and low-energy excitations~\cite{Kringhoj_PRApplied2021, Sagi_2024NatComm, SolePRL_2018}.
Improving hybrid qubit coherence will therefore most likely require engineering cleaner interfaces, suppressing subgap conductance, and incorporating quasiparticle-management strategies~\cite{Bargerbos_PRApplied2023, McEwen_PRL2024}, potentially using larger-gap superconductors.
The co-integrated gatemon–transmon platform introduced here provides a blueprint for quantitatively benchmarking junction-intrinsic dissipation in future hybrid superconducting qubits.

\begin{acknowledgments}
We acknowledge helpful discussions with Karsten Flensberg, Benjamin Levitan, and Alexandre Blais.
We also gratefully acknowledge nanowire growth from Peter Krogstrup and fabrication assistance from Sangeeth Kallatt and Smitha Nair Themadath.

This research was supported by the Novo Nordisk Foundation (grant no. NNF22SA0081175), the NNF Quantum Computing Programme (NQCP), Villum Foundation through a Villum Young Investigator grant (grant no. 37467), the Innovation Fund Denmark (grant no. 2081-00013B, DanQ), the U.S. Army Research Office (grant no. W911NF-22-1-0042, NHyDTech), by the European Union through an ERC Starting Grant, (grant no. 101077479, NovADePro), and by the Carlsberg Foundation (grant no. CF21-0343). 
Any opinions, findings, conclusions or recommendations expressed in this material are those of the author(s) and do not necessarily reflect the views of Army Research Office or the US Government. 
Views and opinions expressed are those of the author(s) only and do not necessarily reflect those of the European Union or the European Research Council. Neither the European Union nor the granting authority can be held responsible for them. 
Finally, we gratefully acknowledge Lena Jacobsen for program management support.
\end{acknowledgments}

\appendix

\section{$T_1$ of the optimized transmons}

\begin{figure}[h]
    \centering
    \includegraphics[width=\columnwidth]{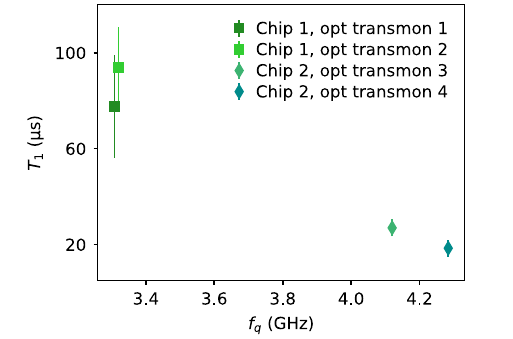}
    \caption{\textbf{Relaxation times of optimized transmons.}
    Two transmons with an optimized design on Chip 1 (marked as squares) are measured to have $T_1=\SI{77.6}{\micro\second}$ and $T_1=\SI{93.8}{\micro\second}$ at $f_q=\SI{3.308}{\giga\hertz}$ and $f_q=\SI{3.320}{\giga\hertz}$, respectively. 
    Two transmons with an optimized design on Chip 2 (marked as diamonds) are measured to have $T_1=\SI{26.9}{\micro\second}$ and $T_1=\SI{18.4}{\micro\second}$ at $f_q=\SI{4.12}{GHz}$ and $f_q=\SI{4.28}{\giga\hertz}$, respectively.
    }
    \label{fig:transmon_T1_supp}
\end{figure}

In our experiment, we also include transmons with a standard design, in which the coupling capacitance of the gate (drive) line and qubit is $\SI{0.01}{\femto\farad}$. 
As shown in Fig.~\ref{fig:transmon_T1_supp}, the two transmons on Chip 1 exhibit $T_1=\SI{77.6}{\micro\second}$ and $T_1=\SI{93.8}{\micro\second}$ at $f_q=\SI{3.308}{\giga\hertz}$ and $f_q=\SI{3.320}{\giga\hertz}$, respectively. 
Two transmons on Chip 2 are measured to have $T_1=\SI{26.9}{\micro\second}$ and $T_1=\SI{18.4}{\micro\second}$ at $f_q=\SI{4.12}{GHz}$ and $f_q=\SI{4.28}{\giga\hertz}$, respectively.

\section{$T_1$ temperature dependence of other devices}

\begin{figure}[h]
    \centering
    \includegraphics[width=\columnwidth]{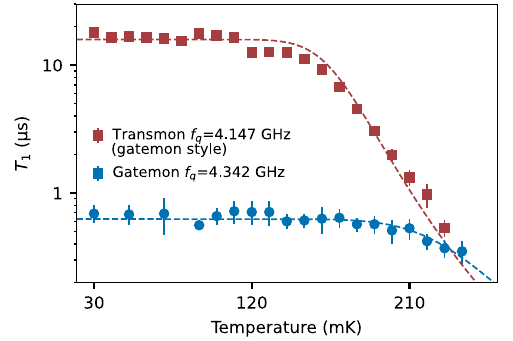}
    \caption{\textbf{Qubit relaxation times $T_1$ as a function of temperature.}
    $T_1$ measurements for another gatemon (blue) and another transmon (red) from Chip 2 at different temperatures. 
    At base temperature, the transmon shows $T_1 = \SI{17.9}{\micro\second}$ at $f_q = \SI{4.147}{\giga\hertz}$, while the gatemon shows $T_1 = \SI{0.69}{\micro\second}$ at $f_q = \SI{4.342}{\giga\hertz}$. The dashed lines represent fits to the quasiparticle poisoning model, the same as the one used in the main text.  Fitting the data gives $\Delta/h = \SI{46.7\pm0.4}{\giga\hertz}$ and $\Delta/h = \SI{50.8\pm0.6}{\giga\hertz}$ for the transmon and gatemon, respectively.}
    \label{fig:T1_vs_temp_supp}
\end{figure}

The $T_1$ temperature dependence of another gatemon and another transmon from Chip 2 is measured and plotted in Fig.~\ref{fig:T1_vs_temp_supp}.
The transmon shows $T_1 = \SI{17.9}{\micro\second}$ at $f_q = \SI{4.147}{\giga\hertz}$ and the gatemon shows $T_1 = \SI{0.69}{\micro\second}$ at $f_q = \SI{4.342}{\giga\hertz}$, when measured at base temperature.
Fitting the data to Eq.~\eqref{eq:T1_QP} gives $\Delta/h = \SI{46.7\pm0.4}{\giga\hertz}$ and $\Delta/h = \SI{50.8\pm0.6}{\giga\hertz}$ for the transmon and gatemon, respectively.
Those parameters fall into the same range as the devices in the main text.

\section{Fabrication recipe}
\label{app:C}
The high-resistivity silicon wafer was first dipped into hydrofluoric acid (HF) to remove native oxide before being loaded into the Plassys MEB550S. 
Titanium is deposited with substrate shutter closed to get better vaccum in the chamber (this step is implemented before all the aluminum evaporations).
A thin film of 200~$\textrm{nm}$ Al was subsequently evaporated onto the silicon substrate at a rate of $\SI{0.2}{\nano\meter/\second}$.
The wafer was then diced into $\SI{1}{\centi\meter}\times\SI{1}{\centi\meter}$ chips.
The control circuitry was defined on the deposited Al film using electron-beam lithography (EBL), followed by a wet-etching process utilizing a Transene aluminum etchant D solution.
The narrow part ($\SI{200}{\nano\meter}$) of the gate line was deposited with a lift-off procedure.
The gate dielectric $\mathrm{HfO_x}$ of $\SI{15}{\nano\meter}$ was grown with atomic layer deposition (ALD) on top of the gate line.

After the fabrication of the circuit, the Al/$\mathrm{AlO_x}$/Al SIS Josephson junctions were fabricated with a standard Manhattan-style double-angle evaporation process for the transmons.
Here, a double-layer resist stack consisting of MMA EL 13 (bottom layer) and CSAR 9 (top layer) was used to provide sufficient stack thickness and resist undercut.
The first junction electrode, consisting of \SI{35}{\nano\meter} of aluminum, was oxidized \textit{in situ} at a pressure of \SI{120}{\milli\bar} for six minutes in the Plassys system, followed by the evaporation of a \SI{125}{\nano\meter} aluminum layer to form the second junction electrode.
The InAs/Al full-shell nanowires were deterministically placed on the chip with a micro-manipulator.
Both the nanowires and the SIS Josephson junction were galvanically connected to the rest of the circuitry via argon ion milling followed by Al evaporation.
The S–Sm–S Josephson junctions were subsequently formed by wet etching away a nominal $\SI{200}{nm}$ opening in the Al shell on the nanowire.

\section{Fridge wiring}
\label{app:D}
\begin{figure}[H]
    \centering
    \includegraphics[width=\columnwidth]{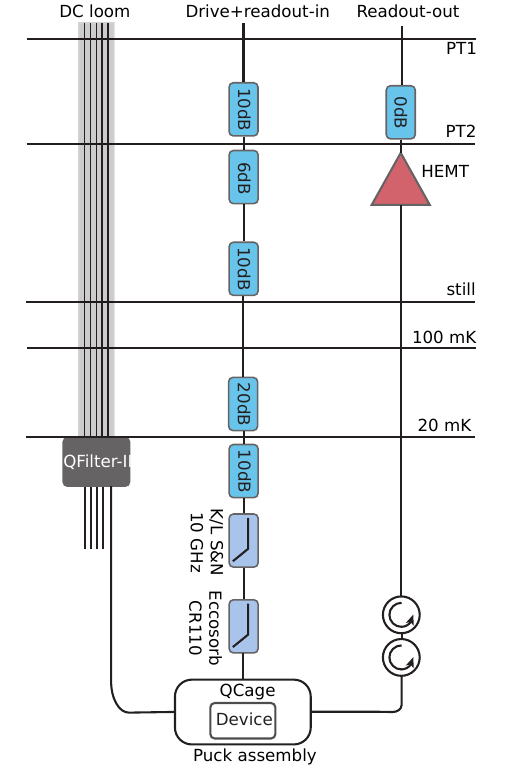}
    \caption{\textbf{Dilution fridge wiring diagram.}
    The DC signal (gate voltage) is filtered by QFilter-II compact low-pass filter (\SI{65}{\kilo\hertz}) and connected to the gatemon gate line. The qubit drive and readout signals are attenuated by attenuators (blue) and filterd by low-pass filters (light blue) before reaching to the feedline on the chip. The outcoming readout signal goes through two isolators and is amplified by high electron mobility transistor (HEMT, red). The chip is wire-bonded and mounted inside the QCage with aluminum shielding.}
    \label{fig:wiring}
\end{figure}

Measurements were carried out in an Oxford Instruments Triton cryofree dilution refrigerator.
The dilution fridge wiring used to measure Chip 3 and Chip 4 is the same as in this reference~\cite{David_25PRApplied}. The fridge wiring diagram used to measure Chip 1 and Chip 2 is shown in Fig.~\ref{fig:wiring}. The DC gate voltage is filterd by QFilter-II low-pass filter with cutoff frequency $\SI{65}{\kilo\hertz}$. The qubit drive signal and readout signal are combined with a microwave splitter at room temperature. The two signals are attenuated and filtered in the fridge before reaching the feedline on the chip. The outcoming readout signal goes through two isolators in series and is amplified by high electron mobility transistor (HEMT) before getting out from the fridge. The chips are wire-bonded and mounted inside the QCage. The aluminum shieldings of the QCage can provide magnetic shielding to the qubits.

\nocite{*}
\bibliographystyle{apsrev4-2}
\bibliography{references}

\begin{thebibliography}{66}%
\makeatletter
\providecommand \@ifxundefined [1]{%
 \@ifx{#1\undefined}
}%
\providecommand \@ifnum [1]{%
 \ifnum #1\expandafter \@firstoftwo
 \else \expandafter \@secondoftwo
 \fi
}%
\providecommand \@ifx [1]{%
 \ifx #1\expandafter \@firstoftwo
 \else \expandafter \@secondoftwo
 \fi
}%
\providecommand \natexlab [1]{#1}%
\providecommand \enquote  [1]{``#1''}%
\providecommand \bibnamefont  [1]{#1}%
\providecommand \bibfnamefont [1]{#1}%
\providecommand \citenamefont [1]{#1}%
\providecommand \href@noop [0]{\@secondoftwo}%
\providecommand \href [0]{\begingroup \@sanitize@url \@href}%
\providecommand \@href[1]{\@@startlink{#1}\@@href}%
\providecommand \@@href[1]{\endgroup#1\@@endlink}%
\providecommand \@sanitize@url [0]{\catcode `\\12\catcode `\$12\catcode `\&12\catcode `\#12\catcode `\^12\catcode `\_12\catcode `\%12\relax}%
\providecommand \@@startlink[1]{}%
\providecommand \@@endlink[0]{}%
\providecommand \url  [0]{\begingroup\@sanitize@url \@url }%
\providecommand \@url [1]{\endgroup\@href {#1}{\urlprefix }}%
\providecommand \urlprefix  [0]{URL }%
\providecommand \Eprint [0]{\href }%
\providecommand \doibase [0]{https://doi.org/}%
\providecommand \selectlanguage [0]{\@gobble}%
\providecommand \bibinfo  [0]{\@secondoftwo}%
\providecommand \bibfield  [0]{\@secondoftwo}%
\providecommand \translation [1]{[#1]}%
\providecommand \BibitemOpen [0]{}%
\providecommand \bibitemStop [0]{}%
\providecommand \bibitemNoStop [0]{.\EOS\space}%
\providecommand \EOS [0]{\spacefactor3000\relax}%
\providecommand \BibitemShut  [1]{\csname bibitem#1\endcsname}%
\let\auto@bib@innerbib\@empty
\bibitem [{\citenamefont {{Nakamura}}\ \emph {et~al.}(1999)\citenamefont {{Nakamura}}, \citenamefont {{Pashkin}},\ and\ \citenamefont {{Tsai}}}]{nakamura1999CPB}%
  \BibitemOpen
  \bibfield  {author} {\bibinfo {author} {\bibfnamefont {Y.}~\bibnamefont {{Nakamura}}}, \bibinfo {author} {\bibfnamefont {Y.~A.}\ \bibnamefont {{Pashkin}}},\ and\ \bibinfo {author} {\bibfnamefont {J.~S.}\ \bibnamefont {{Tsai}}},\ }\href {https://doi.org/10.1038/19718} {\bibfield  {journal} {\bibinfo  {journal} {Nature}\ }\textbf {\bibinfo {volume} {398}},\ \bibinfo {pages} {786} (\bibinfo {year} {1999})}\BibitemShut {NoStop}%
\bibitem [{\citenamefont {Devoret}\ and\ \citenamefont {Schoelkopf}(2013)}]{Schoelkopf2013science}%
  \BibitemOpen
  \bibfield  {author} {\bibinfo {author} {\bibfnamefont {M.~H.}\ \bibnamefont {Devoret}}\ and\ \bibinfo {author} {\bibfnamefont {R.~J.}\ \bibnamefont {Schoelkopf}},\ }\href {https://doi.org/10.1126/science.1231930} {\bibfield  {journal} {\bibinfo  {journal} {Science}\ }\textbf {\bibinfo {volume} {339}},\ \bibinfo {pages} {1169} (\bibinfo {year} {2013})}\BibitemShut {NoStop}%
\bibitem [{\citenamefont {Kjaergaard}\ \emph {et~al.}(2020)\citenamefont {Kjaergaard}, \citenamefont {Schwartz}, \citenamefont {Braumüller}, \citenamefont {Krantz}, \citenamefont {Wang}, \citenamefont {Gustavsson},\ and\ \citenamefont {Oliver}}]{Kjaergaard_2020ARCMP}%
  \BibitemOpen
  \bibfield  {author} {\bibinfo {author} {\bibfnamefont {M.}~\bibnamefont {Kjaergaard}}, \bibinfo {author} {\bibfnamefont {M.~E.}\ \bibnamefont {Schwartz}}, \bibinfo {author} {\bibfnamefont {J.}~\bibnamefont {Braumüller}}, \bibinfo {author} {\bibfnamefont {P.}~\bibnamefont {Krantz}}, \bibinfo {author} {\bibfnamefont {J.~I.-J.}\ \bibnamefont {Wang}}, \bibinfo {author} {\bibfnamefont {S.}~\bibnamefont {Gustavsson}},\ and\ \bibinfo {author} {\bibfnamefont {W.~D.}\ \bibnamefont {Oliver}},\ }\href {http://dx.doi.org/10.1146/annurev-conmatphys-031119-050605} {\bibfield  {journal} {\bibinfo  {journal} {Annual Review of Condensed Matter Physics}\ }\textbf {\bibinfo {volume} {11}},\ \bibinfo {pages} {369–395} (\bibinfo {year} {2020})}\BibitemShut {NoStop}%
\bibitem [{\citenamefont {Danilin}\ and\ \citenamefont {Weides}(2021)}]{danilin2021quantumsensingsuperconductingcircuits}%
  \BibitemOpen
  \bibfield  {author} {\bibinfo {author} {\bibfnamefont {S.}~\bibnamefont {Danilin}}\ and\ \bibinfo {author} {\bibfnamefont {M.}~\bibnamefont {Weides}},\ }\href {https://arxiv.org/abs/2103.11022} {\bibinfo {title} {Quantum sensing with superconducting circuits}} (\bibinfo {year} {2021})\BibitemShut {NoStop}%
\bibitem [{\citenamefont {Wolski}\ \emph {et~al.}(2020)\citenamefont {Wolski}, \citenamefont {Lachance-Quirion}, \citenamefont {Tabuchi}, \citenamefont {Kono}, \citenamefont {Noguchi}, \citenamefont {Usami},\ and\ \citenamefont {Nakamura}}]{Wolski2020PRL}%
  \BibitemOpen
  \bibfield  {author} {\bibinfo {author} {\bibfnamefont {S.}~\bibnamefont {Wolski}}, \bibinfo {author} {\bibfnamefont {D.}~\bibnamefont {Lachance-Quirion}}, \bibinfo {author} {\bibfnamefont {Y.}~\bibnamefont {Tabuchi}}, \bibinfo {author} {\bibfnamefont {S.}~\bibnamefont {Kono}}, \bibinfo {author} {\bibfnamefont {A.}~\bibnamefont {Noguchi}}, \bibinfo {author} {\bibfnamefont {K.}~\bibnamefont {Usami}},\ and\ \bibinfo {author} {\bibfnamefont {Y.}~\bibnamefont {Nakamura}},\ }\href {https://doi.org/10.1103/PhysRevLett.125.117701} {\bibfield  {journal} {\bibinfo  {journal} {Physical Review Letters}\ }\textbf {\bibinfo {volume} {125}} (\bibinfo {year} {2020})}\BibitemShut {NoStop}%
\bibitem [{\citenamefont {Macklin}\ \emph {et~al.}(2015)\citenamefont {Macklin}, \citenamefont {O’Brien}, \citenamefont {Hover}, \citenamefont {Schwartz}, \citenamefont {Bolkhovsky}, \citenamefont {Zhang}, \citenamefont {Oliver},\ and\ \citenamefont {Siddiqi}}]{Siddiqi2015Science}%
  \BibitemOpen
  \bibfield  {author} {\bibinfo {author} {\bibfnamefont {C.}~\bibnamefont {Macklin}}, \bibinfo {author} {\bibfnamefont {K.}~\bibnamefont {O’Brien}}, \bibinfo {author} {\bibfnamefont {D.}~\bibnamefont {Hover}}, \bibinfo {author} {\bibfnamefont {M.~E.}\ \bibnamefont {Schwartz}}, \bibinfo {author} {\bibfnamefont {V.}~\bibnamefont {Bolkhovsky}}, \bibinfo {author} {\bibfnamefont {X.}~\bibnamefont {Zhang}}, \bibinfo {author} {\bibfnamefont {W.~D.}\ \bibnamefont {Oliver}},\ and\ \bibinfo {author} {\bibfnamefont {I.}~\bibnamefont {Siddiqi}},\ }\href {https://doi.org/10.1126/science.aaa8525} {\bibfield  {journal} {\bibinfo  {journal} {Science}\ }\textbf {\bibinfo {volume} {350}},\ \bibinfo {pages} {307} (\bibinfo {year} {2015})}\BibitemShut {NoStop}%
\bibitem [{\citenamefont {Wang}\ \emph {et~al.}(2025)\citenamefont {Wang}, \citenamefont {Peng}, \citenamefont {Knecht}, \citenamefont {Cunningham}, \citenamefont {Lombo}, \citenamefont {Yen}, \citenamefont {Zaidenberg}, \citenamefont {Gingras}, \citenamefont {Niedzielski}, \citenamefont {Stickler}, \citenamefont {Sliwa}, \citenamefont {Serniak}, \citenamefont {Schwartz}, \citenamefont {Oliver},\ and\ \citenamefont {O'Brien}}]{wang2025highefficiencylowlossfloquetmodetraveling}%
  \BibitemOpen
  \bibfield  {author} {\bibinfo {author} {\bibfnamefont {J.}~\bibnamefont {Wang}}, \bibinfo {author} {\bibfnamefont {K.}~\bibnamefont {Peng}}, \bibinfo {author} {\bibfnamefont {J.~M.}\ \bibnamefont {Knecht}}, \bibinfo {author} {\bibfnamefont {G.~D.}\ \bibnamefont {Cunningham}}, \bibinfo {author} {\bibfnamefont {A.~E.}\ \bibnamefont {Lombo}}, \bibinfo {author} {\bibfnamefont {A.}~\bibnamefont {Yen}}, \bibinfo {author} {\bibfnamefont {D.~A.}\ \bibnamefont {Zaidenberg}}, \bibinfo {author} {\bibfnamefont {M.}~\bibnamefont {Gingras}}, \bibinfo {author} {\bibfnamefont {B.~M.}\ \bibnamefont {Niedzielski}}, \bibinfo {author} {\bibfnamefont {H.}~\bibnamefont {Stickler}}, \bibinfo {author} {\bibfnamefont {K.}~\bibnamefont {Sliwa}}, \bibinfo {author} {\bibfnamefont {K.}~\bibnamefont {Serniak}}, \bibinfo {author} {\bibfnamefont {M.~E.}\ \bibnamefont {Schwartz}}, \bibinfo {author} {\bibfnamefont {W.~D.}\ \bibnamefont {Oliver}},\ and\ \bibinfo {author} {\bibfnamefont {K.~P.}\ \bibnamefont {O'Brien}},\ }\href
  {https://arxiv.org/abs/2503.11812} {\bibfield  {journal} {\bibinfo  {journal} {arXiv:2503.11812}\ } (\bibinfo {year} {2025})}\BibitemShut {NoStop}%
\bibitem [{\citenamefont {Chapman}\ \emph {et~al.}(2017)\citenamefont {Chapman}, \citenamefont {Rosenthal}, \citenamefont {Kerckhoff}, \citenamefont {Moores}, \citenamefont {Vale}, \citenamefont {Mates}, \citenamefont {Hilton}, \citenamefont {Lalumi\`ere}, \citenamefont {Blais},\ and\ \citenamefont {Lehnert}}]{Lehnert2017PRX}%
  \BibitemOpen
  \bibfield  {author} {\bibinfo {author} {\bibfnamefont {B.~J.}\ \bibnamefont {Chapman}}, \bibinfo {author} {\bibfnamefont {E.~I.}\ \bibnamefont {Rosenthal}}, \bibinfo {author} {\bibfnamefont {J.}~\bibnamefont {Kerckhoff}}, \bibinfo {author} {\bibfnamefont {B.~A.}\ \bibnamefont {Moores}}, \bibinfo {author} {\bibfnamefont {L.~R.}\ \bibnamefont {Vale}}, \bibinfo {author} {\bibfnamefont {J.~A.~B.}\ \bibnamefont {Mates}}, \bibinfo {author} {\bibfnamefont {G.~C.}\ \bibnamefont {Hilton}}, \bibinfo {author} {\bibfnamefont {K.}~\bibnamefont {Lalumi\`ere}}, \bibinfo {author} {\bibfnamefont {A.}~\bibnamefont {Blais}},\ and\ \bibinfo {author} {\bibfnamefont {K.~W.}\ \bibnamefont {Lehnert}},\ }\href {https://link.aps.org/doi/10.1103/PhysRevX.7.041043} {\bibfield  {journal} {\bibinfo  {journal} {Phys. Rev. X}\ }\textbf {\bibinfo {volume} {7}},\ \bibinfo {pages} {041043} (\bibinfo {year} {2017})}\BibitemShut {NoStop}%
\bibitem [{\citenamefont {White}\ \emph {et~al.}(2015)\citenamefont {White}, \citenamefont {Mutus}, \citenamefont {Hoi}, \citenamefont {Barends}, \citenamefont {Campbell}, \citenamefont {Chen}, \citenamefont {Chen}, \citenamefont {Chiaro}, \citenamefont {Dunsworth}, \citenamefont {Jeffrey}, \citenamefont {Kelly}, \citenamefont {Megrant}, \citenamefont {Neill}, \citenamefont {O'Malley}, \citenamefont {Roushan}, \citenamefont {Sank}, \citenamefont {Vainsencher}, \citenamefont {Wenner}, \citenamefont {Chaudhuri}, \citenamefont {Gao},\ and\ \citenamefont {Martinis}}]{white2015travelingwaveparametricamplifier}%
  \BibitemOpen
  \bibfield  {author} {\bibinfo {author} {\bibfnamefont {T.~C.}\ \bibnamefont {White}}, \bibinfo {author} {\bibfnamefont {J.~Y.}\ \bibnamefont {Mutus}}, \bibinfo {author} {\bibfnamefont {I.-C.}\ \bibnamefont {Hoi}}, \bibinfo {author} {\bibfnamefont {R.}~\bibnamefont {Barends}}, \bibinfo {author} {\bibfnamefont {B.}~\bibnamefont {Campbell}}, \bibinfo {author} {\bibfnamefont {Y.}~\bibnamefont {Chen}}, \bibinfo {author} {\bibfnamefont {Z.}~\bibnamefont {Chen}}, \bibinfo {author} {\bibfnamefont {B.}~\bibnamefont {Chiaro}}, \bibinfo {author} {\bibfnamefont {A.}~\bibnamefont {Dunsworth}}, \bibinfo {author} {\bibfnamefont {E.}~\bibnamefont {Jeffrey}}, \bibinfo {author} {\bibfnamefont {J.}~\bibnamefont {Kelly}}, \bibinfo {author} {\bibfnamefont {A.}~\bibnamefont {Megrant}}, \bibinfo {author} {\bibfnamefont {C.}~\bibnamefont {Neill}}, \bibinfo {author} {\bibfnamefont {P.~J.~J.}\ \bibnamefont {O'Malley}}, \bibinfo {author} {\bibfnamefont {P.}~\bibnamefont {Roushan}}, \bibinfo {author} {\bibfnamefont {D.}~\bibnamefont
  {Sank}}, \bibinfo {author} {\bibfnamefont {A.}~\bibnamefont {Vainsencher}}, \bibinfo {author} {\bibfnamefont {J.}~\bibnamefont {Wenner}}, \bibinfo {author} {\bibfnamefont {S.}~\bibnamefont {Chaudhuri}}, \bibinfo {author} {\bibfnamefont {J.}~\bibnamefont {Gao}},\ and\ \bibinfo {author} {\bibfnamefont {J.~M.}\ \bibnamefont {Martinis}},\ }\href {https://doi.org/10.1063/1.4922348} {\bibfield  {journal} {\bibinfo  {journal} {Applied Physics Letters}\ }\textbf {\bibinfo {volume} {106}},\ \bibinfo {pages} {242601} (\bibinfo {year} {2015})}\BibitemShut {NoStop}%
\bibitem [{\citenamefont {Bland}\ \emph {et~al.}(2025)\citenamefont {Bland}, \citenamefont {Bahrami}, \citenamefont {Martinez}, \citenamefont {Prestegaard}, \citenamefont {Smitham}, \citenamefont {Joshi}, \citenamefont {Hedrick}, \citenamefont {Pakpour-Tabrizi}, \citenamefont {Kumar}, \citenamefont {Jindal}, \citenamefont {Chang}, \citenamefont {Yang}, \citenamefont {Cheng}, \citenamefont {Yao}, \citenamefont {Cava}, \citenamefont {de~Leon},\ and\ \citenamefont {Houck}}]{bland2025Nat}%
  \BibitemOpen
  \bibfield  {author} {\bibinfo {author} {\bibfnamefont {M.~P.}\ \bibnamefont {Bland}}, \bibinfo {author} {\bibfnamefont {F.}~\bibnamefont {Bahrami}}, \bibinfo {author} {\bibfnamefont {J.~G.~C.}\ \bibnamefont {Martinez}}, \bibinfo {author} {\bibfnamefont {P.~H.}\ \bibnamefont {Prestegaard}}, \bibinfo {author} {\bibfnamefont {B.~M.}\ \bibnamefont {Smitham}}, \bibinfo {author} {\bibfnamefont {A.}~\bibnamefont {Joshi}}, \bibinfo {author} {\bibfnamefont {E.}~\bibnamefont {Hedrick}}, \bibinfo {author} {\bibfnamefont {A.}~\bibnamefont {Pakpour-Tabrizi}}, \bibinfo {author} {\bibfnamefont {S.}~\bibnamefont {Kumar}}, \bibinfo {author} {\bibfnamefont {A.}~\bibnamefont {Jindal}}, \bibinfo {author} {\bibfnamefont {R.~D.}\ \bibnamefont {Chang}}, \bibinfo {author} {\bibfnamefont {A.}~\bibnamefont {Yang}}, \bibinfo {author} {\bibfnamefont {G.}~\bibnamefont {Cheng}}, \bibinfo {author} {\bibfnamefont {N.}~\bibnamefont {Yao}}, \bibinfo {author} {\bibfnamefont {R.~J.}\ \bibnamefont {Cava}}, \bibinfo {author} {\bibfnamefont
  {N.~P.}\ \bibnamefont {de~Leon}},\ and\ \bibinfo {author} {\bibfnamefont {A.~A.}\ \bibnamefont {Houck}},\ }\href {https://doi.org/10.1038/s41586-025-09687-4} {\bibfield  {journal} {\bibinfo  {journal} {Nature}\ }\textbf {\bibinfo {volume} {647}},\ \bibinfo {pages} {343} (\bibinfo {year} {2025})}\BibitemShut {NoStop}%
\bibitem [{\citenamefont {Biznárová}\ \emph {et~al.}(2024)\citenamefont {Biznárová}, \citenamefont {Osman}, \citenamefont {Rehnman}, \citenamefont {Chayanun}, \citenamefont {Križan}, \citenamefont {Malmberg}, \citenamefont {Rommel}, \citenamefont {Warren}, \citenamefont {Delsing}, \citenamefont {Yurgens}, \citenamefont {Bylander},\ and\ \citenamefont {Fadavi~Roudsari}}]{Bizn_rov__2024NPJ}%
  \BibitemOpen
  \bibfield  {author} {\bibinfo {author} {\bibfnamefont {J.}~\bibnamefont {Biznárová}}, \bibinfo {author} {\bibfnamefont {A.}~\bibnamefont {Osman}}, \bibinfo {author} {\bibfnamefont {E.}~\bibnamefont {Rehnman}}, \bibinfo {author} {\bibfnamefont {L.}~\bibnamefont {Chayanun}}, \bibinfo {author} {\bibfnamefont {C.}~\bibnamefont {Križan}}, \bibinfo {author} {\bibfnamefont {P.}~\bibnamefont {Malmberg}}, \bibinfo {author} {\bibfnamefont {M.}~\bibnamefont {Rommel}}, \bibinfo {author} {\bibfnamefont {C.}~\bibnamefont {Warren}}, \bibinfo {author} {\bibfnamefont {P.}~\bibnamefont {Delsing}}, \bibinfo {author} {\bibfnamefont {A.}~\bibnamefont {Yurgens}}, \bibinfo {author} {\bibfnamefont {J.}~\bibnamefont {Bylander}},\ and\ \bibinfo {author} {\bibfnamefont {A.}~\bibnamefont {Fadavi~Roudsari}},\ }\href {http://dx.doi.org/10.1038/s41534-024-00868-z} {\bibfield  {journal} {\bibinfo  {journal} {npj Quantum Information}\ }\textbf {\bibinfo {volume} {10}} (\bibinfo {year} {2024})}\BibitemShut {NoStop}%
\bibitem [{\citenamefont {Sivak}\ \emph {et~al.}(2023)\citenamefont {Sivak}, \citenamefont {Eickbusch}, \citenamefont {Royer}, \citenamefont {Singh}, \citenamefont {Tsioutsios}, \citenamefont {Ganjam}, \citenamefont {Miano}, \citenamefont {Brock}, \citenamefont {Ding}, \citenamefont {Frunzio}, \citenamefont {Girvin}, \citenamefont {Schoelkopf},\ and\ \citenamefont {Devoret}}]{Sivak_2023Nat}%
  \BibitemOpen
  \bibfield  {author} {\bibinfo {author} {\bibfnamefont {V.~V.}\ \bibnamefont {Sivak}}, \bibinfo {author} {\bibfnamefont {A.}~\bibnamefont {Eickbusch}}, \bibinfo {author} {\bibfnamefont {B.}~\bibnamefont {Royer}}, \bibinfo {author} {\bibfnamefont {S.}~\bibnamefont {Singh}}, \bibinfo {author} {\bibfnamefont {I.}~\bibnamefont {Tsioutsios}}, \bibinfo {author} {\bibfnamefont {S.}~\bibnamefont {Ganjam}}, \bibinfo {author} {\bibfnamefont {A.}~\bibnamefont {Miano}}, \bibinfo {author} {\bibfnamefont {B.~L.}\ \bibnamefont {Brock}}, \bibinfo {author} {\bibfnamefont {A.~Z.}\ \bibnamefont {Ding}}, \bibinfo {author} {\bibfnamefont {L.}~\bibnamefont {Frunzio}}, \bibinfo {author} {\bibfnamefont {S.~M.}\ \bibnamefont {Girvin}}, \bibinfo {author} {\bibfnamefont {R.~J.}\ \bibnamefont {Schoelkopf}},\ and\ \bibinfo {author} {\bibfnamefont {M.~H.}\ \bibnamefont {Devoret}},\ }\href {http://dx.doi.org/10.1038/s41586-023-05782-6} {\bibfield  {journal} {\bibinfo  {journal} {Nature}\ }\textbf {\bibinfo {volume} {616}},\ \bibinfo
  {pages} {50–55} (\bibinfo {year} {2023})}\BibitemShut {NoStop}%
\bibitem [{\citenamefont {Place}\ \emph {et~al.}(2021)\citenamefont {Place}, \citenamefont {Rodgers}, \citenamefont {Mundada}, \citenamefont {Smitham}, \citenamefont {Fitzpatrick}, \citenamefont {Leng}, \citenamefont {Premkumar}, \citenamefont {Bryon}, \citenamefont {Vrajitoarea}, \citenamefont {Sussman}, \citenamefont {Cheng}, \citenamefont {Madhavan}, \citenamefont {Babla}, \citenamefont {Le}, \citenamefont {Gang}, \citenamefont {Jäck}, \citenamefont {Gyenis}, \citenamefont {Yao}, \citenamefont {Cava}, \citenamefont {de~Leon},\ and\ \citenamefont {Houck}}]{Place_2021NC}%
  \BibitemOpen
  \bibfield  {author} {\bibinfo {author} {\bibfnamefont {A.~P.~M.}\ \bibnamefont {Place}}, \bibinfo {author} {\bibfnamefont {L.~V.~H.}\ \bibnamefont {Rodgers}}, \bibinfo {author} {\bibfnamefont {P.}~\bibnamefont {Mundada}}, \bibinfo {author} {\bibfnamefont {B.~M.}\ \bibnamefont {Smitham}}, \bibinfo {author} {\bibfnamefont {M.}~\bibnamefont {Fitzpatrick}}, \bibinfo {author} {\bibfnamefont {Z.}~\bibnamefont {Leng}}, \bibinfo {author} {\bibfnamefont {A.}~\bibnamefont {Premkumar}}, \bibinfo {author} {\bibfnamefont {J.}~\bibnamefont {Bryon}}, \bibinfo {author} {\bibfnamefont {A.}~\bibnamefont {Vrajitoarea}}, \bibinfo {author} {\bibfnamefont {S.}~\bibnamefont {Sussman}}, \bibinfo {author} {\bibfnamefont {G.}~\bibnamefont {Cheng}}, \bibinfo {author} {\bibfnamefont {T.}~\bibnamefont {Madhavan}}, \bibinfo {author} {\bibfnamefont {H.~K.}\ \bibnamefont {Babla}}, \bibinfo {author} {\bibfnamefont {X.~H.}\ \bibnamefont {Le}}, \bibinfo {author} {\bibfnamefont {Y.}~\bibnamefont {Gang}}, \bibinfo {author} {\bibfnamefont
  {B.}~\bibnamefont {Jäck}}, \bibinfo {author} {\bibfnamefont {A.}~\bibnamefont {Gyenis}}, \bibinfo {author} {\bibfnamefont {N.}~\bibnamefont {Yao}}, \bibinfo {author} {\bibfnamefont {R.~J.}\ \bibnamefont {Cava}}, \bibinfo {author} {\bibfnamefont {N.~P.}\ \bibnamefont {de~Leon}},\ and\ \bibinfo {author} {\bibfnamefont {A.~A.}\ \bibnamefont {Houck}},\ }\href {http://dx.doi.org/10.1038/s41467-021-22030-5} {\bibfield  {journal} {\bibinfo  {journal} {Nature Communications}\ }\textbf {\bibinfo {volume} {12}} (\bibinfo {year} {2021})}\BibitemShut {NoStop}%
\bibitem [{\citenamefont {Berritta}\ \emph {et~al.}(2025)\citenamefont {Berritta}, \citenamefont {Benestad}, \citenamefont {Krzywda}, \citenamefont {Krause}, \citenamefont {Marciniak}, \citenamefont {Kr\o{}jer}, \citenamefont {Warren}, \citenamefont {Hogedal}, \citenamefont {Nylander}, \citenamefont {Ahmad}, \citenamefont {Osman}, \citenamefont {Bizn\'{a}rov\'{a}}, \citenamefont {Rommel}, \citenamefont {Fadavi~Roudsari}, \citenamefont {Bylander}, \citenamefont {Tancredi}, \citenamefont {Danon}, \citenamefont {Hastrup}, \citenamefont {Kuemmeth},\ and\ \citenamefont {Kjaergaard}}]{berritta_arxiv2025}%
  \BibitemOpen
  \bibfield  {author} {\bibinfo {author} {\bibfnamefont {F.}~\bibnamefont {Berritta}}, \bibinfo {author} {\bibfnamefont {J.}~\bibnamefont {Benestad}}, \bibinfo {author} {\bibfnamefont {J.~A.}\ \bibnamefont {Krzywda}}, \bibinfo {author} {\bibfnamefont {O.}~\bibnamefont {Krause}}, \bibinfo {author} {\bibfnamefont {M.~A.}\ \bibnamefont {Marciniak}}, \bibinfo {author} {\bibfnamefont {S.}~\bibnamefont {Kr\o{}jer}}, \bibinfo {author} {\bibfnamefont {C.~W.}\ \bibnamefont {Warren}}, \bibinfo {author} {\bibfnamefont {E.}~\bibnamefont {Hogedal}}, \bibinfo {author} {\bibfnamefont {A.}~\bibnamefont {Nylander}}, \bibinfo {author} {\bibfnamefont {I.}~\bibnamefont {Ahmad}}, \bibinfo {author} {\bibfnamefont {A.}~\bibnamefont {Osman}}, \bibinfo {author} {\bibfnamefont {J.}~\bibnamefont {Bizn\'{a}rov\'{a}}}, \bibinfo {author} {\bibfnamefont {M.}~\bibnamefont {Rommel}}, \bibinfo {author} {\bibfnamefont {A.}~\bibnamefont {Fadavi~Roudsari}}, \bibinfo {author} {\bibfnamefont {J.}~\bibnamefont {Bylander}}, \bibinfo {author}
  {\bibfnamefont {G.}~\bibnamefont {Tancredi}}, \bibinfo {author} {\bibfnamefont {J.}~\bibnamefont {Danon}}, \bibinfo {author} {\bibfnamefont {J.}~\bibnamefont {Hastrup}}, \bibinfo {author} {\bibfnamefont {F.}~\bibnamefont {Kuemmeth}},\ and\ \bibinfo {author} {\bibfnamefont {M.}~\bibnamefont {Kjaergaard}},\ }\href {https://arxiv.org/abs/2506.09576} {\bibfield  {journal} {\bibinfo  {journal} {arXiv:2506.09576}\ } (\bibinfo {year} {2025})}\BibitemShut {NoStop}%
\bibitem [{\citenamefont {Siddiqi}(2021)}]{Siddiqi_2021Engineering}%
  \BibitemOpen
  \bibfield  {author} {\bibinfo {author} {\bibfnamefont {I.}~\bibnamefont {Siddiqi}},\ }\href {https://doi.org/10.1038/s41578-021-00370-4} {\bibfield  {journal} {\bibinfo  {journal} {Nature Reviews Materials}\ }\textbf {\bibinfo {volume} {6}},\ \bibinfo {pages} {875} (\bibinfo {year} {2021})}\BibitemShut {NoStop}%
\bibitem [{\citenamefont {Wang}\ \emph {et~al.}(2015)\citenamefont {Wang}, \citenamefont {Axline}, \citenamefont {Gao}, \citenamefont {Brecht}, \citenamefont {Chu}, \citenamefont {Frunzio}, \citenamefont {Devoret},\ and\ \citenamefont {Schoelkopf}}]{Wang_2015AIP}%
  \BibitemOpen
  \bibfield  {author} {\bibinfo {author} {\bibfnamefont {C.}~\bibnamefont {Wang}}, \bibinfo {author} {\bibfnamefont {C.}~\bibnamefont {Axline}}, \bibinfo {author} {\bibfnamefont {Y.~Y.}\ \bibnamefont {Gao}}, \bibinfo {author} {\bibfnamefont {T.}~\bibnamefont {Brecht}}, \bibinfo {author} {\bibfnamefont {Y.}~\bibnamefont {Chu}}, \bibinfo {author} {\bibfnamefont {L.}~\bibnamefont {Frunzio}}, \bibinfo {author} {\bibfnamefont {M.~H.}\ \bibnamefont {Devoret}},\ and\ \bibinfo {author} {\bibfnamefont {R.~J.}\ \bibnamefont {Schoelkopf}},\ }\href {http://dx.doi.org/10.1063/1.4934486} {\bibfield  {journal} {\bibinfo  {journal} {Applied Physics Letters}\ }\textbf {\bibinfo {volume} {107}} (\bibinfo {year} {2015})}\BibitemShut {NoStop}%
\bibitem [{\citenamefont {Larsen}\ \emph {et~al.}(2015)\citenamefont {Larsen}, \citenamefont {Petersson}, \citenamefont {Kuemmeth}, \citenamefont {Jespersen}, \citenamefont {Krogstrup}, \citenamefont {Nygård},\ and\ \citenamefont {Marcus}}]{Larsen_2015}%
  \BibitemOpen
  \bibfield  {author} {\bibinfo {author} {\bibfnamefont {T.}~\bibnamefont {Larsen}}, \bibinfo {author} {\bibfnamefont {K.}~\bibnamefont {Petersson}}, \bibinfo {author} {\bibfnamefont {F.}~\bibnamefont {Kuemmeth}}, \bibinfo {author} {\bibfnamefont {T.}~\bibnamefont {Jespersen}}, \bibinfo {author} {\bibfnamefont {P.}~\bibnamefont {Krogstrup}}, \bibinfo {author} {\bibfnamefont {J.}~\bibnamefont {Nygård}},\ and\ \bibinfo {author} {\bibfnamefont {C.}~\bibnamefont {Marcus}},\ }\href {http://dx.doi.org/10.1103/PhysRevLett.115.127001} {\bibfield  {journal} {\bibinfo  {journal} {Physical Review Letters}\ }\textbf {\bibinfo {volume} {115}} (\bibinfo {year} {2015})}\BibitemShut {NoStop}%
\bibitem [{\citenamefont {de~Lange}\ \emph {et~al.}(2015)\citenamefont {de~Lange}, \citenamefont {van Heck}, \citenamefont {Bruno}, \citenamefont {van Woerkom}, \citenamefont {Geresdi}, \citenamefont {Plissard}, \citenamefont {Bakkers}, \citenamefont {Akhmerov},\ and\ \citenamefont {DiCarlo}}]{Lange_PRL2015}%
  \BibitemOpen
  \bibfield  {author} {\bibinfo {author} {\bibfnamefont {G.}~\bibnamefont {de~Lange}}, \bibinfo {author} {\bibfnamefont {B.}~\bibnamefont {van Heck}}, \bibinfo {author} {\bibfnamefont {A.}~\bibnamefont {Bruno}}, \bibinfo {author} {\bibfnamefont {D.~J.}\ \bibnamefont {van Woerkom}}, \bibinfo {author} {\bibfnamefont {A.}~\bibnamefont {Geresdi}}, \bibinfo {author} {\bibfnamefont {S.~R.}\ \bibnamefont {Plissard}}, \bibinfo {author} {\bibfnamefont {E.~P. A.~M.}\ \bibnamefont {Bakkers}}, \bibinfo {author} {\bibfnamefont {A.~R.}\ \bibnamefont {Akhmerov}},\ and\ \bibinfo {author} {\bibfnamefont {L.}~\bibnamefont {DiCarlo}},\ }\href {https://link.aps.org/doi/10.1103/PhysRevLett.115.127002} {\bibfield  {journal} {\bibinfo  {journal} {Phys. Rev. Lett.}\ }\textbf {\bibinfo {volume} {115}},\ \bibinfo {pages} {127002} (\bibinfo {year} {2015})}\BibitemShut {NoStop}%
\bibitem [{\citenamefont {Casparis}\ \emph {et~al.}(2018)\citenamefont {Casparis}, \citenamefont {Connolly}, \citenamefont {Kjaergaard}, \citenamefont {Pearson}, \citenamefont {Kringhøj}, \citenamefont {Larsen}, \citenamefont {Kuemmeth}, \citenamefont {Wang}, \citenamefont {Thomas}, \citenamefont {Gronin}, \citenamefont {Gardner}, \citenamefont {Manfra}, \citenamefont {Marcus},\ and\ \citenamefont {Petersson}}]{Casparis_2018}%
  \BibitemOpen
  \bibfield  {author} {\bibinfo {author} {\bibfnamefont {L.}~\bibnamefont {Casparis}}, \bibinfo {author} {\bibfnamefont {M.~R.}\ \bibnamefont {Connolly}}, \bibinfo {author} {\bibfnamefont {M.}~\bibnamefont {Kjaergaard}}, \bibinfo {author} {\bibfnamefont {N.~J.}\ \bibnamefont {Pearson}}, \bibinfo {author} {\bibfnamefont {A.}~\bibnamefont {Kringhøj}}, \bibinfo {author} {\bibfnamefont {T.~W.}\ \bibnamefont {Larsen}}, \bibinfo {author} {\bibfnamefont {F.}~\bibnamefont {Kuemmeth}}, \bibinfo {author} {\bibfnamefont {T.}~\bibnamefont {Wang}}, \bibinfo {author} {\bibfnamefont {C.}~\bibnamefont {Thomas}}, \bibinfo {author} {\bibfnamefont {S.}~\bibnamefont {Gronin}}, \bibinfo {author} {\bibfnamefont {G.~C.}\ \bibnamefont {Gardner}}, \bibinfo {author} {\bibfnamefont {M.~J.}\ \bibnamefont {Manfra}}, \bibinfo {author} {\bibfnamefont {C.~M.}\ \bibnamefont {Marcus}},\ and\ \bibinfo {author} {\bibfnamefont {K.~D.}\ \bibnamefont {Petersson}},\ }\href {http://dx.doi.org/10.1038/s41565-018-0207-y} {\bibfield  {journal}
  {\bibinfo  {journal} {Nature Nanotechnology}\ }\textbf {\bibinfo {volume} {13}},\ \bibinfo {pages} {915–919} (\bibinfo {year} {2018})}\BibitemShut {NoStop}%
\bibitem [{\citenamefont {Strickland}\ \emph {et~al.}(2025)\citenamefont {Strickland}, \citenamefont {Elfeky}, \citenamefont {Baker}, \citenamefont {Maiani}, \citenamefont {Lee}, \citenamefont {Levy}, \citenamefont {Issokson}, \citenamefont {Vrajitoarea},\ and\ \citenamefont {Shabani}}]{Strickland_2025}%
  \BibitemOpen
  \bibfield  {author} {\bibinfo {author} {\bibfnamefont {W.~M.}\ \bibnamefont {Strickland}}, \bibinfo {author} {\bibfnamefont {B.~H.}\ \bibnamefont {Elfeky}}, \bibinfo {author} {\bibfnamefont {L.}~\bibnamefont {Baker}}, \bibinfo {author} {\bibfnamefont {A.}~\bibnamefont {Maiani}}, \bibinfo {author} {\bibfnamefont {J.}~\bibnamefont {Lee}}, \bibinfo {author} {\bibfnamefont {I.}~\bibnamefont {Levy}}, \bibinfo {author} {\bibfnamefont {J.}~\bibnamefont {Issokson}}, \bibinfo {author} {\bibfnamefont {A.}~\bibnamefont {Vrajitoarea}},\ and\ \bibinfo {author} {\bibfnamefont {J.}~\bibnamefont {Shabani}},\ }\href {http://dx.doi.org/10.1103/PRXQuantum.6.010326} {\bibfield  {journal} {\bibinfo  {journal} {PRX Quantum}\ }\textbf {\bibinfo {volume} {6}} (\bibinfo {year} {2025})}\BibitemShut {NoStop}%
\bibitem [{\citenamefont {Hays}\ \emph {et~al.}(2021)\citenamefont {Hays}, \citenamefont {Fatemi}, \citenamefont {Bouman}, \citenamefont {Cerrillo}, \citenamefont {Diamond}, \citenamefont {Serniak}, \citenamefont {Connolly}, \citenamefont {Krogstrup}, \citenamefont {Nygård}, \citenamefont {Levy~Yeyati}, \citenamefont {Geresdi},\ and\ \citenamefont {Devoret}}]{Hays_2021}%
  \BibitemOpen
  \bibfield  {author} {\bibinfo {author} {\bibfnamefont {M.}~\bibnamefont {Hays}}, \bibinfo {author} {\bibfnamefont {V.}~\bibnamefont {Fatemi}}, \bibinfo {author} {\bibfnamefont {D.}~\bibnamefont {Bouman}}, \bibinfo {author} {\bibfnamefont {J.}~\bibnamefont {Cerrillo}}, \bibinfo {author} {\bibfnamefont {S.}~\bibnamefont {Diamond}}, \bibinfo {author} {\bibfnamefont {K.}~\bibnamefont {Serniak}}, \bibinfo {author} {\bibfnamefont {T.}~\bibnamefont {Connolly}}, \bibinfo {author} {\bibfnamefont {P.}~\bibnamefont {Krogstrup}}, \bibinfo {author} {\bibfnamefont {J.}~\bibnamefont {Nygård}}, \bibinfo {author} {\bibfnamefont {A.}~\bibnamefont {Levy~Yeyati}}, \bibinfo {author} {\bibfnamefont {A.}~\bibnamefont {Geresdi}},\ and\ \bibinfo {author} {\bibfnamefont {M.~H.}\ \bibnamefont {Devoret}},\ }\href {http://dx.doi.org/10.1126/science.abf0345} {\bibfield  {journal} {\bibinfo  {journal} {Science}\ }\textbf {\bibinfo {volume} {373}},\ \bibinfo {pages} {430–433} (\bibinfo {year} {2021})}\BibitemShut {NoStop}%
\bibitem [{\citenamefont {Pita-Vidal}\ \emph {et~al.}(2023)\citenamefont {Pita-Vidal}, \citenamefont {Bargerbos}, \citenamefont {Žitko}, \citenamefont {Splitthoff}, \citenamefont {Grünhaupt}, \citenamefont {Wesdorp}, \citenamefont {Liu}, \citenamefont {Kouwenhoven}, \citenamefont {Aguado}, \citenamefont {van Heck}, \citenamefont {Kou},\ and\ \citenamefont {Andersen}}]{Pita_Vidal_2023}%
  \BibitemOpen
  \bibfield  {author} {\bibinfo {author} {\bibfnamefont {M.}~\bibnamefont {Pita-Vidal}}, \bibinfo {author} {\bibfnamefont {A.}~\bibnamefont {Bargerbos}}, \bibinfo {author} {\bibfnamefont {R.}~\bibnamefont {Žitko}}, \bibinfo {author} {\bibfnamefont {L.~J.}\ \bibnamefont {Splitthoff}}, \bibinfo {author} {\bibfnamefont {L.}~\bibnamefont {Grünhaupt}}, \bibinfo {author} {\bibfnamefont {J.~J.}\ \bibnamefont {Wesdorp}}, \bibinfo {author} {\bibfnamefont {Y.}~\bibnamefont {Liu}}, \bibinfo {author} {\bibfnamefont {L.~P.}\ \bibnamefont {Kouwenhoven}}, \bibinfo {author} {\bibfnamefont {R.}~\bibnamefont {Aguado}}, \bibinfo {author} {\bibfnamefont {B.}~\bibnamefont {van Heck}}, \bibinfo {author} {\bibfnamefont {A.}~\bibnamefont {Kou}},\ and\ \bibinfo {author} {\bibfnamefont {C.~K.}\ \bibnamefont {Andersen}},\ }\href {http://dx.doi.org/10.1038/s41567-023-02071-x} {\bibfield  {journal} {\bibinfo  {journal} {Nature Physics}\ }\textbf {\bibinfo {volume} {19}},\ \bibinfo {pages} {1110–1115} (\bibinfo {year}
  {2023})}\BibitemShut {NoStop}%
\bibitem [{\citenamefont {Luthi}\ \emph {et~al.}(2018)\citenamefont {Luthi}, \citenamefont {Stavenga}, \citenamefont {Enzing}, \citenamefont {Bruno}, \citenamefont {Dickel}, \citenamefont {Langford}, \citenamefont {Rol}, \citenamefont {Jespersen}, \citenamefont {Nyg\aa{}rd}, \citenamefont {Krogstrup},\ and\ \citenamefont {DiCarlo}}]{Luthi_PRL2018}%
  \BibitemOpen
  \bibfield  {author} {\bibinfo {author} {\bibfnamefont {F.}~\bibnamefont {Luthi}}, \bibinfo {author} {\bibfnamefont {T.}~\bibnamefont {Stavenga}}, \bibinfo {author} {\bibfnamefont {O.~W.}\ \bibnamefont {Enzing}}, \bibinfo {author} {\bibfnamefont {A.}~\bibnamefont {Bruno}}, \bibinfo {author} {\bibfnamefont {C.}~\bibnamefont {Dickel}}, \bibinfo {author} {\bibfnamefont {N.~K.}\ \bibnamefont {Langford}}, \bibinfo {author} {\bibfnamefont {M.~A.}\ \bibnamefont {Rol}}, \bibinfo {author} {\bibfnamefont {T.~S.}\ \bibnamefont {Jespersen}}, \bibinfo {author} {\bibfnamefont {J.}~\bibnamefont {Nyg\aa{}rd}}, \bibinfo {author} {\bibfnamefont {P.}~\bibnamefont {Krogstrup}},\ and\ \bibinfo {author} {\bibfnamefont {L.}~\bibnamefont {DiCarlo}},\ }\href {https://link.aps.org/doi/10.1103/PhysRevLett.120.100502} {\bibfield  {journal} {\bibinfo  {journal} {Phys. Rev. Lett.}\ }\textbf {\bibinfo {volume} {120}},\ \bibinfo {pages} {100502} (\bibinfo {year} {2018})}\BibitemShut {NoStop}%
\bibitem [{\citenamefont {Purkayastha}\ \emph {et~al.}(2025)\citenamefont {Purkayastha}, \citenamefont {Sharma}, \citenamefont {Patel}, \citenamefont {Chen}, \citenamefont {Dempsey}, \citenamefont {Asodekar}, \citenamefont {Sinha}, \citenamefont {Tomasian}, \citenamefont {Pendharkar}, \citenamefont {Palmstr{\o}m}, \citenamefont {Hocevar}, \citenamefont {Zuo}, \citenamefont {Hatridge},\ and\ \citenamefont {Frolov}}]{purkayastha_arxiv2025}%
  \BibitemOpen
  \bibfield  {author} {\bibinfo {author} {\bibfnamefont {A.}~\bibnamefont {Purkayastha}}, \bibinfo {author} {\bibfnamefont {A.}~\bibnamefont {Sharma}}, \bibinfo {author} {\bibfnamefont {P.~J.}\ \bibnamefont {Patel}}, \bibinfo {author} {\bibfnamefont {A.-H.}\ \bibnamefont {Chen}}, \bibinfo {author} {\bibfnamefont {C.~P.}\ \bibnamefont {Dempsey}}, \bibinfo {author} {\bibfnamefont {S.}~\bibnamefont {Asodekar}}, \bibinfo {author} {\bibfnamefont {S.}~\bibnamefont {Sinha}}, \bibinfo {author} {\bibfnamefont {M.}~\bibnamefont {Tomasian}}, \bibinfo {author} {\bibfnamefont {M.}~\bibnamefont {Pendharkar}}, \bibinfo {author} {\bibfnamefont {C.~J.}\ \bibnamefont {Palmstr{\o}m}}, \bibinfo {author} {\bibfnamefont {M.}~\bibnamefont {Hocevar}}, \bibinfo {author} {\bibfnamefont {K.}~\bibnamefont {Zuo}}, \bibinfo {author} {\bibfnamefont {M.}~\bibnamefont {Hatridge}},\ and\ \bibinfo {author} {\bibfnamefont {S.~M.}\ \bibnamefont {Frolov}},\ }\href {https://arxiv.org/abs/2508.04007} {\bibfield  {journal} {\bibinfo  {journal}
  {arXiv:2508.04007}\ } (\bibinfo {year} {2025})}\BibitemShut {NoStop}%
\bibitem [{\citenamefont {Casparis}\ \emph {et~al.}(2016)\citenamefont {Casparis}, \citenamefont {Larsen}, \citenamefont {Olsen}, \citenamefont {Kuemmeth}, \citenamefont {Krogstrup}, \citenamefont {Nyg\aa{}rd}, \citenamefont {Petersson},\ and\ \citenamefont {Marcus}}]{Casparis_PRL2016}%
  \BibitemOpen
  \bibfield  {author} {\bibinfo {author} {\bibfnamefont {L.}~\bibnamefont {Casparis}}, \bibinfo {author} {\bibfnamefont {T.~W.}\ \bibnamefont {Larsen}}, \bibinfo {author} {\bibfnamefont {M.~S.}\ \bibnamefont {Olsen}}, \bibinfo {author} {\bibfnamefont {F.}~\bibnamefont {Kuemmeth}}, \bibinfo {author} {\bibfnamefont {P.}~\bibnamefont {Krogstrup}}, \bibinfo {author} {\bibfnamefont {J.}~\bibnamefont {Nyg\aa{}rd}}, \bibinfo {author} {\bibfnamefont {K.~D.}\ \bibnamefont {Petersson}},\ and\ \bibinfo {author} {\bibfnamefont {C.~M.}\ \bibnamefont {Marcus}},\ }\href {https://link.aps.org/doi/10.1103/PhysRevLett.116.150505} {\bibfield  {journal} {\bibinfo  {journal} {Phys. Rev. Lett.}\ }\textbf {\bibinfo {volume} {116}},\ \bibinfo {pages} {150505} (\bibinfo {year} {2016})}\BibitemShut {NoStop}%
\bibitem [{\citenamefont {Kringhøj}\ \emph {et~al.}(2021)\citenamefont {Kringhøj}, \citenamefont {Larsen}, \citenamefont {Erlandsson}, \citenamefont {Uilhoorn}, \citenamefont {Kroll}, \citenamefont {Hesselberg}, \citenamefont {McNeil}, \citenamefont {Krogstrup}, \citenamefont {Casparis}, \citenamefont {Marcus},\ and\ \citenamefont {Petersson}}]{Kringhoj_PRApplied2021}%
  \BibitemOpen
  \bibfield  {author} {\bibinfo {author} {\bibfnamefont {A.}~\bibnamefont {Kringhøj}}, \bibinfo {author} {\bibfnamefont {T.~W.}\ \bibnamefont {Larsen}}, \bibinfo {author} {\bibfnamefont {O.}~\bibnamefont {Erlandsson}}, \bibinfo {author} {\bibfnamefont {W.}~\bibnamefont {Uilhoorn}}, \bibinfo {author} {\bibfnamefont {J.}~\bibnamefont {Kroll}}, \bibinfo {author} {\bibfnamefont {M.}~\bibnamefont {Hesselberg}}, \bibinfo {author} {\bibfnamefont {R.}~\bibnamefont {McNeil}}, \bibinfo {author} {\bibfnamefont {P.}~\bibnamefont {Krogstrup}}, \bibinfo {author} {\bibfnamefont {L.}~\bibnamefont {Casparis}}, \bibinfo {author} {\bibfnamefont {C.}~\bibnamefont {Marcus}},\ and\ \bibinfo {author} {\bibfnamefont {K.}~\bibnamefont {Petersson}},\ }\href {http://dx.doi.org/10.1103/PhysRevApplied.15.054001} {\bibfield  {journal} {\bibinfo  {journal} {Physical Review Applied}\ }\textbf {\bibinfo {volume} {15}} (\bibinfo {year} {2021})}\BibitemShut {NoStop}%
\bibitem [{\citenamefont {Hertel}\ \emph {et~al.}(2022)\citenamefont {Hertel}, \citenamefont {Eichinger}, \citenamefont {Andersen}, \citenamefont {van Zanten}, \citenamefont {Kallatt}, \citenamefont {Scarlino}, \citenamefont {Kringhøj}, \citenamefont {Chavez-Garcia}, \citenamefont {Gardner}, \citenamefont {Gronin}, \citenamefont {Manfra}, \citenamefont {Gyenis}, \citenamefont {Kjaergaard}, \citenamefont {Marcus},\ and\ \citenamefont {Petersson}}]{Hertel_2022PRApplied}%
  \BibitemOpen
  \bibfield  {author} {\bibinfo {author} {\bibfnamefont {A.}~\bibnamefont {Hertel}}, \bibinfo {author} {\bibfnamefont {M.}~\bibnamefont {Eichinger}}, \bibinfo {author} {\bibfnamefont {L.~O.}\ \bibnamefont {Andersen}}, \bibinfo {author} {\bibfnamefont {D.~M.}\ \bibnamefont {van Zanten}}, \bibinfo {author} {\bibfnamefont {S.}~\bibnamefont {Kallatt}}, \bibinfo {author} {\bibfnamefont {P.}~\bibnamefont {Scarlino}}, \bibinfo {author} {\bibfnamefont {A.}~\bibnamefont {Kringhøj}}, \bibinfo {author} {\bibfnamefont {J.~M.}\ \bibnamefont {Chavez-Garcia}}, \bibinfo {author} {\bibfnamefont {G.~C.}\ \bibnamefont {Gardner}}, \bibinfo {author} {\bibfnamefont {S.}~\bibnamefont {Gronin}}, \bibinfo {author} {\bibfnamefont {M.~J.}\ \bibnamefont {Manfra}}, \bibinfo {author} {\bibfnamefont {A.}~\bibnamefont {Gyenis}}, \bibinfo {author} {\bibfnamefont {M.}~\bibnamefont {Kjaergaard}}, \bibinfo {author} {\bibfnamefont {C.~M.}\ \bibnamefont {Marcus}},\ and\ \bibinfo {author} {\bibfnamefont {K.~D.}\ \bibnamefont {Petersson}},\ }\href
  {http://dx.doi.org/10.1103/PhysRevApplied.18.034042} {\bibfield  {journal} {\bibinfo  {journal} {Physical Review Applied}\ }\textbf {\bibinfo {volume} {18}} (\bibinfo {year} {2022})}\BibitemShut {NoStop}%
\bibitem [{\citenamefont {Sagi}\ \emph {et~al.}(2024)\citenamefont {Sagi}, \citenamefont {Crippa}, \citenamefont {Valentini}, \citenamefont {Janik}, \citenamefont {Baghumyan}, \citenamefont {Fabris}, \citenamefont {Kapoor}, \citenamefont {Hassani}, \citenamefont {Fink}, \citenamefont {Calcaterra}, \citenamefont {Chrastina}, \citenamefont {Isella},\ and\ \citenamefont {Katsaros}}]{Sagi_2024NatComm}%
  \BibitemOpen
  \bibfield  {author} {\bibinfo {author} {\bibfnamefont {O.}~\bibnamefont {Sagi}}, \bibinfo {author} {\bibfnamefont {A.}~\bibnamefont {Crippa}}, \bibinfo {author} {\bibfnamefont {M.}~\bibnamefont {Valentini}}, \bibinfo {author} {\bibfnamefont {M.}~\bibnamefont {Janik}}, \bibinfo {author} {\bibfnamefont {L.}~\bibnamefont {Baghumyan}}, \bibinfo {author} {\bibfnamefont {G.}~\bibnamefont {Fabris}}, \bibinfo {author} {\bibfnamefont {L.}~\bibnamefont {Kapoor}}, \bibinfo {author} {\bibfnamefont {F.}~\bibnamefont {Hassani}}, \bibinfo {author} {\bibfnamefont {J.}~\bibnamefont {Fink}}, \bibinfo {author} {\bibfnamefont {S.}~\bibnamefont {Calcaterra}}, \bibinfo {author} {\bibfnamefont {D.}~\bibnamefont {Chrastina}}, \bibinfo {author} {\bibfnamefont {G.}~\bibnamefont {Isella}},\ and\ \bibinfo {author} {\bibfnamefont {G.}~\bibnamefont {Katsaros}},\ }\href {http://dx.doi.org/10.1038/s41467-024-50763-6} {\bibfield  {journal} {\bibinfo  {journal} {Nature Communications}\ }\textbf {\bibinfo {volume} {15}} (\bibinfo {year}
  {2024})}\BibitemShut {NoStop}%
\bibitem [{\citenamefont {Riechert}\ \emph {et~al.}(2025)\citenamefont {Riechert}, \citenamefont {Annabi}, \citenamefont {Peugeot}, \citenamefont {Duprez}, \citenamefont {Hantute}, \citenamefont {Watanabe}, \citenamefont {Taniguchi}, \citenamefont {Arrighi}, \citenamefont {Griesmar}, \citenamefont {Pillet},\ and\ \citenamefont {Bretheau}}]{Riechert_2025NatComm}%
  \BibitemOpen
  \bibfield  {author} {\bibinfo {author} {\bibfnamefont {H.}~\bibnamefont {Riechert}}, \bibinfo {author} {\bibfnamefont {S.}~\bibnamefont {Annabi}}, \bibinfo {author} {\bibfnamefont {A.}~\bibnamefont {Peugeot}}, \bibinfo {author} {\bibfnamefont {H.}~\bibnamefont {Duprez}}, \bibinfo {author} {\bibfnamefont {M.}~\bibnamefont {Hantute}}, \bibinfo {author} {\bibfnamefont {K.}~\bibnamefont {Watanabe}}, \bibinfo {author} {\bibfnamefont {T.}~\bibnamefont {Taniguchi}}, \bibinfo {author} {\bibfnamefont {E.}~\bibnamefont {Arrighi}}, \bibinfo {author} {\bibfnamefont {J.}~\bibnamefont {Griesmar}}, \bibinfo {author} {\bibfnamefont {J.-D.}\ \bibnamefont {Pillet}},\ and\ \bibinfo {author} {\bibfnamefont {L.}~\bibnamefont {Bretheau}},\ }\href {http://dx.doi.org/10.1038/s41467-025-62283-y} {\bibfield  {journal} {\bibinfo  {journal} {Nature Communications}\ }\textbf {\bibinfo {volume} {16}} (\bibinfo {year} {2025})}\BibitemShut {NoStop}%
\bibitem [{\citenamefont {Zheng}\ \emph {et~al.}(2024)\citenamefont {Zheng}, \citenamefont {Cheung}, \citenamefont {Sangwan}, \citenamefont {Kononov}, \citenamefont {Haller}, \citenamefont {Ridderbos}, \citenamefont {Ciaccia}, \citenamefont {Ungerer}, \citenamefont {Li}, \citenamefont {Bakkers}, \citenamefont {Baumgartner},\ and\ \citenamefont {Sch{\"o}nenberger}}]{Zheng_NanoLett2024}%
  \BibitemOpen
  \bibfield  {author} {\bibinfo {author} {\bibfnamefont {H.}~\bibnamefont {Zheng}}, \bibinfo {author} {\bibfnamefont {L.~Y.}\ \bibnamefont {Cheung}}, \bibinfo {author} {\bibfnamefont {N.}~\bibnamefont {Sangwan}}, \bibinfo {author} {\bibfnamefont {A.}~\bibnamefont {Kononov}}, \bibinfo {author} {\bibfnamefont {R.}~\bibnamefont {Haller}}, \bibinfo {author} {\bibfnamefont {J.}~\bibnamefont {Ridderbos}}, \bibinfo {author} {\bibfnamefont {C.}~\bibnamefont {Ciaccia}}, \bibinfo {author} {\bibfnamefont {J.~H.}\ \bibnamefont {Ungerer}}, \bibinfo {author} {\bibfnamefont {A.}~\bibnamefont {Li}}, \bibinfo {author} {\bibfnamefont {E.~P.}\ \bibnamefont {Bakkers}}, \bibinfo {author} {\bibfnamefont {A.}~\bibnamefont {Baumgartner}},\ and\ \bibinfo {author} {\bibfnamefont {C.}~\bibnamefont {Sch{\"o}nenberger}},\ }\href {https://doi.org/10.1021/acs.nanolett.4c00770} {\bibfield  {journal} {\bibinfo  {journal} {Nano Letters}\ }\textbf {\bibinfo {volume} {24}},\ \bibinfo {pages} {7173} (\bibinfo {year} {2024})}\BibitemShut
  {NoStop}%
\bibitem [{\citenamefont {Feldstein-Bofill}\ \emph {et~al.}(2025)\citenamefont {Feldstein-Bofill}, \citenamefont {Sun}, \citenamefont {Wied}, \citenamefont {Singh}, \citenamefont {Isakov}, \citenamefont {Kr\o{}jer}, \citenamefont {Hastrup}, \citenamefont {Gyenis},\ and\ \citenamefont {Kjaergaard}}]{David_25PRApplied}%
  \BibitemOpen
  \bibfield  {author} {\bibinfo {author} {\bibfnamefont {D.}~\bibnamefont {Feldstein-Bofill}}, \bibinfo {author} {\bibfnamefont {Z.}~\bibnamefont {Sun}}, \bibinfo {author} {\bibfnamefont {C.}~\bibnamefont {Wied}}, \bibinfo {author} {\bibfnamefont {S.}~\bibnamefont {Singh}}, \bibinfo {author} {\bibfnamefont {B.~D.}\ \bibnamefont {Isakov}}, \bibinfo {author} {\bibfnamefont {S.}~\bibnamefont {Kr\o{}jer}}, \bibinfo {author} {\bibfnamefont {J.}~\bibnamefont {Hastrup}}, \bibinfo {author} {\bibfnamefont {A.}~\bibnamefont {Gyenis}},\ and\ \bibinfo {author} {\bibfnamefont {M.}~\bibnamefont {Kjaergaard}},\ }\href {https://link.aps.org/doi/10.1103/d68y-sqzm} {\bibfield  {journal} {\bibinfo  {journal} {Phys. Rev. Appl.}\ }\textbf {\bibinfo {volume} {24}},\ \bibinfo {pages} {044099} (\bibinfo {year} {2025})}\BibitemShut {NoStop}%
\bibitem [{\citenamefont {Strickland}\ \emph {et~al.}(2024)\citenamefont {Strickland}, \citenamefont {Baker}, \citenamefont {Lee}, \citenamefont {Dindial}, \citenamefont {Elfeky}, \citenamefont {Strohbeen}, \citenamefont {Hatefipour}, \citenamefont {Yu}, \citenamefont {Levy}, \citenamefont {Issokson}, \citenamefont {Manucharyan},\ and\ \citenamefont {Shabani}}]{Strickland_PRR2024}%
  \BibitemOpen
  \bibfield  {author} {\bibinfo {author} {\bibfnamefont {W.~M.}\ \bibnamefont {Strickland}}, \bibinfo {author} {\bibfnamefont {L.~J.}\ \bibnamefont {Baker}}, \bibinfo {author} {\bibfnamefont {J.}~\bibnamefont {Lee}}, \bibinfo {author} {\bibfnamefont {K.}~\bibnamefont {Dindial}}, \bibinfo {author} {\bibfnamefont {B.~H.}\ \bibnamefont {Elfeky}}, \bibinfo {author} {\bibfnamefont {P.~J.}\ \bibnamefont {Strohbeen}}, \bibinfo {author} {\bibfnamefont {M.}~\bibnamefont {Hatefipour}}, \bibinfo {author} {\bibfnamefont {P.}~\bibnamefont {Yu}}, \bibinfo {author} {\bibfnamefont {I.}~\bibnamefont {Levy}}, \bibinfo {author} {\bibfnamefont {J.}~\bibnamefont {Issokson}}, \bibinfo {author} {\bibfnamefont {V.~E.}\ \bibnamefont {Manucharyan}},\ and\ \bibinfo {author} {\bibfnamefont {J.}~\bibnamefont {Shabani}},\ }\href {https://link.aps.org/doi/10.1103/PhysRevResearch.6.023094} {\bibfield  {journal} {\bibinfo  {journal} {Phys. Rev. Res.}\ }\textbf {\bibinfo {volume} {6}},\ \bibinfo {pages} {023094} (\bibinfo {year}
  {2024})}\BibitemShut {NoStop}%
\bibitem [{\citenamefont {Houck}\ \emph {et~al.}(2008)\citenamefont {Houck}, \citenamefont {Schreier}, \citenamefont {Johnson}, \citenamefont {Chow}, \citenamefont {Koch}, \citenamefont {Gambetta}, \citenamefont {Schuster}, \citenamefont {Frunzio}, \citenamefont {Devoret}, \citenamefont {Girvin},\ and\ \citenamefont {Schoelkopf}}]{Houck_08PRL}%
  \BibitemOpen
  \bibfield  {author} {\bibinfo {author} {\bibfnamefont {A.~A.}\ \bibnamefont {Houck}}, \bibinfo {author} {\bibfnamefont {J.~A.}\ \bibnamefont {Schreier}}, \bibinfo {author} {\bibfnamefont {B.~R.}\ \bibnamefont {Johnson}}, \bibinfo {author} {\bibfnamefont {J.~M.}\ \bibnamefont {Chow}}, \bibinfo {author} {\bibfnamefont {J.}~\bibnamefont {Koch}}, \bibinfo {author} {\bibfnamefont {J.~M.}\ \bibnamefont {Gambetta}}, \bibinfo {author} {\bibfnamefont {D.~I.}\ \bibnamefont {Schuster}}, \bibinfo {author} {\bibfnamefont {L.}~\bibnamefont {Frunzio}}, \bibinfo {author} {\bibfnamefont {M.~H.}\ \bibnamefont {Devoret}}, \bibinfo {author} {\bibfnamefont {S.~M.}\ \bibnamefont {Girvin}},\ and\ \bibinfo {author} {\bibfnamefont {R.~J.}\ \bibnamefont {Schoelkopf}},\ }\href {https://link.aps.org/doi/10.1103/PhysRevLett.101.080502} {\bibfield  {journal} {\bibinfo  {journal} {Phys. Rev. Lett.}\ }\textbf {\bibinfo {volume} {101}},\ \bibinfo {pages} {080502} (\bibinfo {year} {2008})}\BibitemShut {NoStop}%
\bibitem [{\citenamefont {Catelani}\ \emph {et~al.}(2011)\citenamefont {Catelani}, \citenamefont {Schoelkopf}, \citenamefont {Devoret},\ and\ \citenamefont {Glazman}}]{Catelani_2011PRB}%
  \BibitemOpen
  \bibfield  {author} {\bibinfo {author} {\bibfnamefont {G.}~\bibnamefont {Catelani}}, \bibinfo {author} {\bibfnamefont {R.~J.}\ \bibnamefont {Schoelkopf}}, \bibinfo {author} {\bibfnamefont {M.~H.}\ \bibnamefont {Devoret}},\ and\ \bibinfo {author} {\bibfnamefont {L.~I.}\ \bibnamefont {Glazman}},\ }\href {http://dx.doi.org/10.1103/PhysRevB.84.064517} {\bibfield  {journal} {\bibinfo  {journal} {Physical Review B}\ }\textbf {\bibinfo {volume} {84}} (\bibinfo {year} {2011})}\BibitemShut {NoStop}%
\bibitem [{\citenamefont {Catelani}\ \emph {et~al.}(2012)\citenamefont {Catelani}, \citenamefont {Nigg}, \citenamefont {Girvin}, \citenamefont {Schoelkopf},\ and\ \citenamefont {Glazman}}]{Catelani_12PRB}%
  \BibitemOpen
  \bibfield  {author} {\bibinfo {author} {\bibfnamefont {G.}~\bibnamefont {Catelani}}, \bibinfo {author} {\bibfnamefont {S.~E.}\ \bibnamefont {Nigg}}, \bibinfo {author} {\bibfnamefont {S.~M.}\ \bibnamefont {Girvin}}, \bibinfo {author} {\bibfnamefont {R.~J.}\ \bibnamefont {Schoelkopf}},\ and\ \bibinfo {author} {\bibfnamefont {L.~I.}\ \bibnamefont {Glazman}},\ }\href {https://doi.org/10.1103/PhysRevB.86.184514} {\bibfield  {journal} {\bibinfo  {journal} {Physical Review B}\ }\textbf {\bibinfo {volume} {86}},\ \bibinfo {pages} {184514} (\bibinfo {year} {2012})}\BibitemShut {NoStop}%
\bibitem [{\citenamefont {Serniak}\ \emph {et~al.}(2018)\citenamefont {Serniak}, \citenamefont {Hays}, \citenamefont {de~Lange}, \citenamefont {Diamond}, \citenamefont {Shankar}, \citenamefont {Burkhart}, \citenamefont {Frunzio}, \citenamefont {Houzet},\ and\ \citenamefont {Devoret}}]{Kyle_18PRL}%
  \BibitemOpen
  \bibfield  {author} {\bibinfo {author} {\bibfnamefont {K.}~\bibnamefont {Serniak}}, \bibinfo {author} {\bibfnamefont {M.}~\bibnamefont {Hays}}, \bibinfo {author} {\bibfnamefont {G.}~\bibnamefont {de~Lange}}, \bibinfo {author} {\bibfnamefont {S.}~\bibnamefont {Diamond}}, \bibinfo {author} {\bibfnamefont {S.}~\bibnamefont {Shankar}}, \bibinfo {author} {\bibfnamefont {L.~D.}\ \bibnamefont {Burkhart}}, \bibinfo {author} {\bibfnamefont {L.}~\bibnamefont {Frunzio}}, \bibinfo {author} {\bibfnamefont {M.}~\bibnamefont {Houzet}},\ and\ \bibinfo {author} {\bibfnamefont {M.~H.}\ \bibnamefont {Devoret}},\ }\href {https://link.aps.org/doi/10.1103/PhysRevLett.121.157701} {\bibfield  {journal} {\bibinfo  {journal} {Phys. Rev. Lett.}\ }\textbf {\bibinfo {volume} {121}},\ \bibinfo {pages} {157701} (\bibinfo {year} {2018})}\BibitemShut {NoStop}%
\bibitem [{\citenamefont {Chang}\ \emph {et~al.}(2015)\citenamefont {Chang}, \citenamefont {Albrecht}, \citenamefont {Jespersen}, \citenamefont {Kuemmeth}, \citenamefont {Krogstrup}, \citenamefont {Nygård},\ and\ \citenamefont {Marcus}}]{Chang_NatNano2015}%
  \BibitemOpen
  \bibfield  {author} {\bibinfo {author} {\bibfnamefont {W.}~\bibnamefont {Chang}}, \bibinfo {author} {\bibfnamefont {S.~M.}\ \bibnamefont {Albrecht}}, \bibinfo {author} {\bibfnamefont {T.~S.}\ \bibnamefont {Jespersen}}, \bibinfo {author} {\bibfnamefont {F.}~\bibnamefont {Kuemmeth}}, \bibinfo {author} {\bibfnamefont {P.}~\bibnamefont {Krogstrup}}, \bibinfo {author} {\bibfnamefont {J.}~\bibnamefont {Nygård}},\ and\ \bibinfo {author} {\bibfnamefont {C.~M.}\ \bibnamefont {Marcus}},\ }\href {http://dx.doi.org/10.1038/nnano.2014.306} {\bibfield  {journal} {\bibinfo  {journal} {Nature Nanotechnology}\ }\textbf {\bibinfo {volume} {10}},\ \bibinfo {pages} {232–236} (\bibinfo {year} {2015})}\BibitemShut {NoStop}%
\bibitem [{\citenamefont {Valentini}\ \emph {et~al.}(2024)\citenamefont {Valentini}, \citenamefont {Sagi}, \citenamefont {Baghumyan}, \citenamefont {de~Gijsel}, \citenamefont {Jung}, \citenamefont {Calcaterra}, \citenamefont {Ballabio}, \citenamefont {Aguilera~Servin}, \citenamefont {Aggarwal}, \citenamefont {Janik}, \citenamefont {Adletzberger}, \citenamefont {Seoane~Souto}, \citenamefont {Leijnse}, \citenamefont {Danon}, \citenamefont {Schrade}, \citenamefont {Bakkers}, \citenamefont {Chrastina}, \citenamefont {Isella},\ and\ \citenamefont {Katsaros}}]{Valentini_NatComm2024}%
  \BibitemOpen
  \bibfield  {author} {\bibinfo {author} {\bibfnamefont {M.}~\bibnamefont {Valentini}}, \bibinfo {author} {\bibfnamefont {O.}~\bibnamefont {Sagi}}, \bibinfo {author} {\bibfnamefont {L.}~\bibnamefont {Baghumyan}}, \bibinfo {author} {\bibfnamefont {T.}~\bibnamefont {de~Gijsel}}, \bibinfo {author} {\bibfnamefont {J.}~\bibnamefont {Jung}}, \bibinfo {author} {\bibfnamefont {S.}~\bibnamefont {Calcaterra}}, \bibinfo {author} {\bibfnamefont {A.}~\bibnamefont {Ballabio}}, \bibinfo {author} {\bibfnamefont {J.}~\bibnamefont {Aguilera~Servin}}, \bibinfo {author} {\bibfnamefont {K.}~\bibnamefont {Aggarwal}}, \bibinfo {author} {\bibfnamefont {M.}~\bibnamefont {Janik}}, \bibinfo {author} {\bibfnamefont {T.}~\bibnamefont {Adletzberger}}, \bibinfo {author} {\bibfnamefont {R.}~\bibnamefont {Seoane~Souto}}, \bibinfo {author} {\bibfnamefont {M.}~\bibnamefont {Leijnse}}, \bibinfo {author} {\bibfnamefont {J.}~\bibnamefont {Danon}}, \bibinfo {author} {\bibfnamefont {C.}~\bibnamefont {Schrade}}, \bibinfo {author} {\bibfnamefont
  {E.}~\bibnamefont {Bakkers}}, \bibinfo {author} {\bibfnamefont {D.}~\bibnamefont {Chrastina}}, \bibinfo {author} {\bibfnamefont {G.}~\bibnamefont {Isella}},\ and\ \bibinfo {author} {\bibfnamefont {G.}~\bibnamefont {Katsaros}},\ }\href {http://dx.doi.org/10.1038/s41467-023-44114-0} {\bibfield  {journal} {\bibinfo  {journal} {Nature Communications}\ }\textbf {\bibinfo {volume} {15}} (\bibinfo {year} {2024})}\BibitemShut {NoStop}%
\bibitem [{\citenamefont {Liu}\ \emph {et~al.}(2017)\citenamefont {Liu}, \citenamefont {Setiawan}, \citenamefont {Sau},\ and\ \citenamefont {Das~Sarma}}]{Liu_PRB2017}%
  \BibitemOpen
  \bibfield  {author} {\bibinfo {author} {\bibfnamefont {C.-X.}\ \bibnamefont {Liu}}, \bibinfo {author} {\bibfnamefont {F.}~\bibnamefont {Setiawan}}, \bibinfo {author} {\bibfnamefont {J.~D.}\ \bibnamefont {Sau}},\ and\ \bibinfo {author} {\bibfnamefont {S.}~\bibnamefont {Das~Sarma}},\ }\href {https://link.aps.org/doi/10.1103/PhysRevB.96.054520} {\bibfield  {journal} {\bibinfo  {journal} {Phys. Rev. B}\ }\textbf {\bibinfo {volume} {96}},\ \bibinfo {pages} {054520} (\bibinfo {year} {2017})}\BibitemShut {NoStop}%
\bibitem [{\citenamefont {Takei}\ \emph {et~al.}(2013)\citenamefont {Takei}, \citenamefont {Fregoso}, \citenamefont {Hui}, \citenamefont {Lobos},\ and\ \citenamefont {Das~Sarma}}]{Sarma_PRL2013}%
  \BibitemOpen
  \bibfield  {author} {\bibinfo {author} {\bibfnamefont {S.}~\bibnamefont {Takei}}, \bibinfo {author} {\bibfnamefont {B.~M.}\ \bibnamefont {Fregoso}}, \bibinfo {author} {\bibfnamefont {H.-Y.}\ \bibnamefont {Hui}}, \bibinfo {author} {\bibfnamefont {A.~M.}\ \bibnamefont {Lobos}},\ and\ \bibinfo {author} {\bibfnamefont {S.}~\bibnamefont {Das~Sarma}},\ }\href {https://link.aps.org/doi/10.1103/PhysRevLett.110.186803} {\bibfield  {journal} {\bibinfo  {journal} {Phys. Rev. Lett.}\ }\textbf {\bibinfo {volume} {110}},\ \bibinfo {pages} {186803} (\bibinfo {year} {2013})}\BibitemShut {NoStop}%
\bibitem [{\citenamefont {Levajac}\ \emph {et~al.}(2023)\citenamefont {Levajac}, \citenamefont {Wang}, \citenamefont {Sfiligoj} \emph {et~al.}}]{Levajac_NatComm2023}%
  \BibitemOpen
  \bibfield  {author} {\bibinfo {author} {\bibfnamefont {V.}~\bibnamefont {Levajac}}, \bibinfo {author} {\bibfnamefont {J.~Y.}\ \bibnamefont {Wang}}, \bibinfo {author} {\bibfnamefont {C.}~\bibnamefont {Sfiligoj}}, \emph {et~al.},\ }\href {https://doi.org/10.1038/s41467-023-42422-z} {\bibfield  {journal} {\bibinfo  {journal} {Nature Communications}\ }\textbf {\bibinfo {volume} {14}},\ \bibinfo {pages} {6647} (\bibinfo {year} {2023})}\BibitemShut {NoStop}%
\bibitem [{\citenamefont {Manenti}\ \emph {et~al.}(2021)\citenamefont {Manenti}, \citenamefont {Sete}, \citenamefont {Chen}, \citenamefont {Kulshreshtha}, \citenamefont {Yeh}, \citenamefont {Oruc}, \citenamefont {Bestwick}, \citenamefont {Field}, \citenamefont {Jackson},\ and\ \citenamefont {Poletto}}]{Manenti_APL2021}%
  \BibitemOpen
  \bibfield  {author} {\bibinfo {author} {\bibfnamefont {R.}~\bibnamefont {Manenti}}, \bibinfo {author} {\bibfnamefont {E.~A.}\ \bibnamefont {Sete}}, \bibinfo {author} {\bibfnamefont {A.~Q.}\ \bibnamefont {Chen}}, \bibinfo {author} {\bibfnamefont {S.}~\bibnamefont {Kulshreshtha}}, \bibinfo {author} {\bibfnamefont {J.-H.}\ \bibnamefont {Yeh}}, \bibinfo {author} {\bibfnamefont {F.}~\bibnamefont {Oruc}}, \bibinfo {author} {\bibfnamefont {A.}~\bibnamefont {Bestwick}}, \bibinfo {author} {\bibfnamefont {M.}~\bibnamefont {Field}}, \bibinfo {author} {\bibfnamefont {K.}~\bibnamefont {Jackson}},\ and\ \bibinfo {author} {\bibfnamefont {S.}~\bibnamefont {Poletto}},\ }\href {http://dx.doi.org/10.1063/5.0065517} {\bibfield  {journal} {\bibinfo  {journal} {Applied Physics Letters}\ }\textbf {\bibinfo {volume} {119}} (\bibinfo {year} {2021})}\BibitemShut {NoStop}%
\bibitem [{\citenamefont {Kringh\o{}j}\ \emph {et~al.}(2018)\citenamefont {Kringh\o{}j}, \citenamefont {Casparis}, \citenamefont {Hell}, \citenamefont {Larsen}, \citenamefont {Kuemmeth}, \citenamefont {Leijnse}, \citenamefont {Flensberg}, \citenamefont {Krogstrup}, \citenamefont {Nyg\aa{}rd}, \citenamefont {Petersson},\ and\ \citenamefont {Marcus}}]{Anders_PRB2018}%
  \BibitemOpen
  \bibfield  {author} {\bibinfo {author} {\bibfnamefont {A.}~\bibnamefont {Kringh\o{}j}}, \bibinfo {author} {\bibfnamefont {L.}~\bibnamefont {Casparis}}, \bibinfo {author} {\bibfnamefont {M.}~\bibnamefont {Hell}}, \bibinfo {author} {\bibfnamefont {T.~W.}\ \bibnamefont {Larsen}}, \bibinfo {author} {\bibfnamefont {F.}~\bibnamefont {Kuemmeth}}, \bibinfo {author} {\bibfnamefont {M.}~\bibnamefont {Leijnse}}, \bibinfo {author} {\bibfnamefont {K.}~\bibnamefont {Flensberg}}, \bibinfo {author} {\bibfnamefont {P.}~\bibnamefont {Krogstrup}}, \bibinfo {author} {\bibfnamefont {J.}~\bibnamefont {Nyg\aa{}rd}}, \bibinfo {author} {\bibfnamefont {K.~D.}\ \bibnamefont {Petersson}},\ and\ \bibinfo {author} {\bibfnamefont {C.~M.}\ \bibnamefont {Marcus}},\ }\href {https://link.aps.org/doi/10.1103/PhysRevB.97.060508} {\bibfield  {journal} {\bibinfo  {journal} {Phys. Rev. B}\ }\textbf {\bibinfo {volume} {97}},\ \bibinfo {pages} {060508} (\bibinfo {year} {2018})}\BibitemShut {NoStop}%
\bibitem [{\citenamefont {Bruno}\ \emph {et~al.}(2015)\citenamefont {Bruno}, \citenamefont {de~Lange}, \citenamefont {Asaad}, \citenamefont {van~der Enden}, \citenamefont {Langford},\ and\ \citenamefont {DiCarlo}}]{Bruno_APL2015}%
  \BibitemOpen
  \bibfield  {author} {\bibinfo {author} {\bibfnamefont {A.}~\bibnamefont {Bruno}}, \bibinfo {author} {\bibfnamefont {G.}~\bibnamefont {de~Lange}}, \bibinfo {author} {\bibfnamefont {S.}~\bibnamefont {Asaad}}, \bibinfo {author} {\bibfnamefont {K.~L.}\ \bibnamefont {van~der Enden}}, \bibinfo {author} {\bibfnamefont {N.~K.}\ \bibnamefont {Langford}},\ and\ \bibinfo {author} {\bibfnamefont {L.}~\bibnamefont {DiCarlo}},\ }\href {http://dx.doi.org/10.1063/1.4919761} {\bibfield  {journal} {\bibinfo  {journal} {Applied Physics Letters}\ }\textbf {\bibinfo {volume} {106}} (\bibinfo {year} {2015})}\BibitemShut {NoStop}%
\bibitem [{\citenamefont {Murray}(2021)}]{Murray_2021}%
  \BibitemOpen
  \bibfield  {author} {\bibinfo {author} {\bibfnamefont {C.~E.}\ \bibnamefont {Murray}},\ }\href {http://dx.doi.org/10.1016/j.mser.2021.100646} {\bibfield  {journal} {\bibinfo  {journal} {Materials Science and Engineering: R: Reports}\ }\textbf {\bibinfo {volume} {146}},\ \bibinfo {pages} {100646} (\bibinfo {year} {2021})}\BibitemShut {NoStop}%
\bibitem [{\citenamefont {Nersisyan}\ \emph {et~al.}(2019)\citenamefont {Nersisyan}, \citenamefont {Poletto}, \citenamefont {Alidoust}, \citenamefont {Manenti}, \citenamefont {Renzas}, \citenamefont {Bui}, \citenamefont {Vu}, \citenamefont {Whyland}, \citenamefont {Mohan}, \citenamefont {Sete}, \citenamefont {Stanwyck}, \citenamefont {Bestwick},\ and\ \citenamefont {Reagor}}]{Ani_arxiv2019}%
  \BibitemOpen
  \bibfield  {author} {\bibinfo {author} {\bibfnamefont {A.}~\bibnamefont {Nersisyan}}, \bibinfo {author} {\bibfnamefont {S.}~\bibnamefont {Poletto}}, \bibinfo {author} {\bibfnamefont {N.}~\bibnamefont {Alidoust}}, \bibinfo {author} {\bibfnamefont {R.}~\bibnamefont {Manenti}}, \bibinfo {author} {\bibfnamefont {R.}~\bibnamefont {Renzas}}, \bibinfo {author} {\bibfnamefont {C.-V.}\ \bibnamefont {Bui}}, \bibinfo {author} {\bibfnamefont {K.}~\bibnamefont {Vu}}, \bibinfo {author} {\bibfnamefont {T.}~\bibnamefont {Whyland}}, \bibinfo {author} {\bibfnamefont {Y.}~\bibnamefont {Mohan}}, \bibinfo {author} {\bibfnamefont {E.~A.}\ \bibnamefont {Sete}}, \bibinfo {author} {\bibfnamefont {S.}~\bibnamefont {Stanwyck}}, \bibinfo {author} {\bibfnamefont {A.}~\bibnamefont {Bestwick}},\ and\ \bibinfo {author} {\bibfnamefont {M.}~\bibnamefont {Reagor}},\ }\href {https://arxiv.org/abs/1901.08042} {\bibinfo {title} {Manufacturing low dissipation superconducting quantum processors}} (\bibinfo {year} {2019})\BibitemShut {NoStop}%
\bibitem [{\citenamefont {Neeley}\ \emph {et~al.}(2008)\citenamefont {Neeley}, \citenamefont {Ansmann}, \citenamefont {Bialczak}, \citenamefont {Hofheinz}, \citenamefont {Katz}, \citenamefont {Lucero}, \citenamefont {O'Connell}, \citenamefont {Wang}, \citenamefont {Cleland},\ and\ \citenamefont {Martinis}}]{Neely_PhysRevB2008}%
  \BibitemOpen
  \bibfield  {author} {\bibinfo {author} {\bibfnamefont {M.}~\bibnamefont {Neeley}}, \bibinfo {author} {\bibfnamefont {M.}~\bibnamefont {Ansmann}}, \bibinfo {author} {\bibfnamefont {R.~C.}\ \bibnamefont {Bialczak}}, \bibinfo {author} {\bibfnamefont {M.}~\bibnamefont {Hofheinz}}, \bibinfo {author} {\bibfnamefont {N.}~\bibnamefont {Katz}}, \bibinfo {author} {\bibfnamefont {E.}~\bibnamefont {Lucero}}, \bibinfo {author} {\bibfnamefont {A.}~\bibnamefont {O'Connell}}, \bibinfo {author} {\bibfnamefont {H.}~\bibnamefont {Wang}}, \bibinfo {author} {\bibfnamefont {A.~N.}\ \bibnamefont {Cleland}},\ and\ \bibinfo {author} {\bibfnamefont {J.~M.}\ \bibnamefont {Martinis}},\ }\href {https://link.aps.org/doi/10.1103/PhysRevB.77.180508} {\bibfield  {journal} {\bibinfo  {journal} {Phys. Rev. B}\ }\textbf {\bibinfo {volume} {77}},\ \bibinfo {pages} {180508} (\bibinfo {year} {2008})}\BibitemShut {NoStop}%
\bibitem [{\citenamefont {Sun}\ \emph {et~al.}(2012)\citenamefont {Sun}, \citenamefont {DiCarlo}, \citenamefont {Reed}, \citenamefont {Catelani}, \citenamefont {Bishop}, \citenamefont {Schuster}, \citenamefont {Johnson}, \citenamefont {Yang}, \citenamefont {Frunzio}, \citenamefont {Glazman}, \citenamefont {Devoret},\ and\ \citenamefont {Schoelkopf}}]{Sun_PRL2012}%
  \BibitemOpen
  \bibfield  {author} {\bibinfo {author} {\bibfnamefont {L.}~\bibnamefont {Sun}}, \bibinfo {author} {\bibfnamefont {L.}~\bibnamefont {DiCarlo}}, \bibinfo {author} {\bibfnamefont {M.~D.}\ \bibnamefont {Reed}}, \bibinfo {author} {\bibfnamefont {G.}~\bibnamefont {Catelani}}, \bibinfo {author} {\bibfnamefont {L.~S.}\ \bibnamefont {Bishop}}, \bibinfo {author} {\bibfnamefont {D.~I.}\ \bibnamefont {Schuster}}, \bibinfo {author} {\bibfnamefont {B.~R.}\ \bibnamefont {Johnson}}, \bibinfo {author} {\bibfnamefont {G.~A.}\ \bibnamefont {Yang}}, \bibinfo {author} {\bibfnamefont {L.}~\bibnamefont {Frunzio}}, \bibinfo {author} {\bibfnamefont {L.}~\bibnamefont {Glazman}}, \bibinfo {author} {\bibfnamefont {M.~H.}\ \bibnamefont {Devoret}},\ and\ \bibinfo {author} {\bibfnamefont {R.~J.}\ \bibnamefont {Schoelkopf}},\ }\href {https://link.aps.org/doi/10.1103/PhysRevLett.108.230509} {\bibfield  {journal} {\bibinfo  {journal} {Phys. Rev. Lett.}\ }\textbf {\bibinfo {volume} {108}},\ \bibinfo {pages} {230509} (\bibinfo {year}
  {2012})}\BibitemShut {NoStop}%
\bibitem [{\citenamefont {Ristè}\ \emph {et~al.}(2013)\citenamefont {Ristè}, \citenamefont {Bultink}, \citenamefont {Tiggelman}, \citenamefont {Schouten}, \citenamefont {Lehnert},\ and\ \citenamefont {DiCarlo}}]{Rist_NatComm2013}%
  \BibitemOpen
  \bibfield  {author} {\bibinfo {author} {\bibfnamefont {D.}~\bibnamefont {Ristè}}, \bibinfo {author} {\bibfnamefont {C.~C.}\ \bibnamefont {Bultink}}, \bibinfo {author} {\bibfnamefont {M.~J.}\ \bibnamefont {Tiggelman}}, \bibinfo {author} {\bibfnamefont {R.~N.}\ \bibnamefont {Schouten}}, \bibinfo {author} {\bibfnamefont {K.~W.}\ \bibnamefont {Lehnert}},\ and\ \bibinfo {author} {\bibfnamefont {L.}~\bibnamefont {DiCarlo}},\ }\href {http://dx.doi.org/10.1038/ncomms2936} {\bibfield  {journal} {\bibinfo  {journal} {Nature Communications}\ }\textbf {\bibinfo {volume} {4}} (\bibinfo {year} {2013})}\BibitemShut {NoStop}%
\bibitem [{\citenamefont {Krause}\ \emph {et~al.}(2024)\citenamefont {Krause}, \citenamefont {Marchegiani}, \citenamefont {Janssen}, \citenamefont {Catelani}, \citenamefont {Ando},\ and\ \citenamefont {Dickel}}]{Krause_PRApplied2024}%
  \BibitemOpen
  \bibfield  {author} {\bibinfo {author} {\bibfnamefont {J.}~\bibnamefont {Krause}}, \bibinfo {author} {\bibfnamefont {G.}~\bibnamefont {Marchegiani}}, \bibinfo {author} {\bibfnamefont {L.}~\bibnamefont {Janssen}}, \bibinfo {author} {\bibfnamefont {G.}~\bibnamefont {Catelani}}, \bibinfo {author} {\bibfnamefont {Y.}~\bibnamefont {Ando}},\ and\ \bibinfo {author} {\bibfnamefont {C.}~\bibnamefont {Dickel}},\ }\href {https://link.aps.org/doi/10.1103/PhysRevApplied.22.044063} {\bibfield  {journal} {\bibinfo  {journal} {Phys. Rev. Appl.}\ }\textbf {\bibinfo {volume} {22}},\ \bibinfo {pages} {044063} (\bibinfo {year} {2024})}\BibitemShut {NoStop}%
\bibitem [{\citenamefont {Connolly}\ \emph {et~al.}(2024)\citenamefont {Connolly}, \citenamefont {Kurilovich}, \citenamefont {Diamond}, \citenamefont {Nho}, \citenamefont {B\o{}ttcher}, \citenamefont {Glazman}, \citenamefont {Fatemi},\ and\ \citenamefont {Devoret}}]{Connoly_PRL2024}%
  \BibitemOpen
  \bibfield  {author} {\bibinfo {author} {\bibfnamefont {T.}~\bibnamefont {Connolly}}, \bibinfo {author} {\bibfnamefont {P.~D.}\ \bibnamefont {Kurilovich}}, \bibinfo {author} {\bibfnamefont {S.}~\bibnamefont {Diamond}}, \bibinfo {author} {\bibfnamefont {H.}~\bibnamefont {Nho}}, \bibinfo {author} {\bibfnamefont {C.~G.~L.}\ \bibnamefont {B\o{}ttcher}}, \bibinfo {author} {\bibfnamefont {L.~I.}\ \bibnamefont {Glazman}}, \bibinfo {author} {\bibfnamefont {V.}~\bibnamefont {Fatemi}},\ and\ \bibinfo {author} {\bibfnamefont {M.~H.}\ \bibnamefont {Devoret}},\ }\href {https://link.aps.org/doi/10.1103/PhysRevLett.132.217001} {\bibfield  {journal} {\bibinfo  {journal} {Phys. Rev. Lett.}\ }\textbf {\bibinfo {volume} {132}},\ \bibinfo {pages} {217001} (\bibinfo {year} {2024})}\BibitemShut {NoStop}%
\bibitem [{\citenamefont {Diamond}\ \emph {et~al.}(2022)\citenamefont {Diamond}, \citenamefont {Fatemi}, \citenamefont {Hays}, \citenamefont {Nho}, \citenamefont {Kurilovich}, \citenamefont {Connolly}, \citenamefont {Joshi}, \citenamefont {Serniak}, \citenamefont {Frunzio}, \citenamefont {Glazman},\ and\ \citenamefont {Devoret}}]{Diamond_PRXQuantum2022}%
  \BibitemOpen
  \bibfield  {author} {\bibinfo {author} {\bibfnamefont {S.}~\bibnamefont {Diamond}}, \bibinfo {author} {\bibfnamefont {V.}~\bibnamefont {Fatemi}}, \bibinfo {author} {\bibfnamefont {M.}~\bibnamefont {Hays}}, \bibinfo {author} {\bibfnamefont {H.}~\bibnamefont {Nho}}, \bibinfo {author} {\bibfnamefont {P.~D.}\ \bibnamefont {Kurilovich}}, \bibinfo {author} {\bibfnamefont {T.}~\bibnamefont {Connolly}}, \bibinfo {author} {\bibfnamefont {V.~R.}\ \bibnamefont {Joshi}}, \bibinfo {author} {\bibfnamefont {K.}~\bibnamefont {Serniak}}, \bibinfo {author} {\bibfnamefont {L.}~\bibnamefont {Frunzio}}, \bibinfo {author} {\bibfnamefont {L.~I.}\ \bibnamefont {Glazman}},\ and\ \bibinfo {author} {\bibfnamefont {M.~H.}\ \bibnamefont {Devoret}},\ }\href {https://link.aps.org/doi/10.1103/PRXQuantum.3.040304} {\bibfield  {journal} {\bibinfo  {journal} {PRX Quantum}\ }\textbf {\bibinfo {volume} {3}},\ \bibinfo {pages} {040304} (\bibinfo {year} {2022})}\BibitemShut {NoStop}%
\bibitem [{\citenamefont {Liu}\ \emph {et~al.}(2024)\citenamefont {Liu}, \citenamefont {Harrison}, \citenamefont {Patel}, \citenamefont {Wilen}, \citenamefont {Rafferty}, \citenamefont {Shearrow}, \citenamefont {Ballard}, \citenamefont {Iaia}, \citenamefont {Ku}, \citenamefont {Plourde},\ and\ \citenamefont {McDermott}}]{Liu_PRL2024}%
  \BibitemOpen
  \bibfield  {author} {\bibinfo {author} {\bibfnamefont {C.~H.}\ \bibnamefont {Liu}}, \bibinfo {author} {\bibfnamefont {D.~C.}\ \bibnamefont {Harrison}}, \bibinfo {author} {\bibfnamefont {S.}~\bibnamefont {Patel}}, \bibinfo {author} {\bibfnamefont {C.~D.}\ \bibnamefont {Wilen}}, \bibinfo {author} {\bibfnamefont {O.}~\bibnamefont {Rafferty}}, \bibinfo {author} {\bibfnamefont {A.}~\bibnamefont {Shearrow}}, \bibinfo {author} {\bibfnamefont {A.}~\bibnamefont {Ballard}}, \bibinfo {author} {\bibfnamefont {V.}~\bibnamefont {Iaia}}, \bibinfo {author} {\bibfnamefont {J.}~\bibnamefont {Ku}}, \bibinfo {author} {\bibfnamefont {B.~L.~T.}\ \bibnamefont {Plourde}},\ and\ \bibinfo {author} {\bibfnamefont {R.}~\bibnamefont {McDermott}},\ }\href {https://link.aps.org/doi/10.1103/PhysRevLett.132.017001} {\bibfield  {journal} {\bibinfo  {journal} {Phys. Rev. Lett.}\ }\textbf {\bibinfo {volume} {132}},\ \bibinfo {pages} {017001} (\bibinfo {year} {2024})}\BibitemShut {NoStop}%
\bibitem [{\citenamefont {Benevides}\ \emph {et~al.}(2024)\citenamefont {Benevides}, \citenamefont {Drimmer}, \citenamefont {Bisson}, \citenamefont {Adinolfi}, \citenamefont {L\"upke}, \citenamefont {Doeleman}, \citenamefont {Catelani},\ and\ \citenamefont {Chu}}]{Benevides_PRL2024}%
  \BibitemOpen
  \bibfield  {author} {\bibinfo {author} {\bibfnamefont {R.}~\bibnamefont {Benevides}}, \bibinfo {author} {\bibfnamefont {M.}~\bibnamefont {Drimmer}}, \bibinfo {author} {\bibfnamefont {G.}~\bibnamefont {Bisson}}, \bibinfo {author} {\bibfnamefont {F.}~\bibnamefont {Adinolfi}}, \bibinfo {author} {\bibfnamefont {U.~v.}\ \bibnamefont {L\"upke}}, \bibinfo {author} {\bibfnamefont {H.~M.}\ \bibnamefont {Doeleman}}, \bibinfo {author} {\bibfnamefont {G.}~\bibnamefont {Catelani}},\ and\ \bibinfo {author} {\bibfnamefont {Y.}~\bibnamefont {Chu}},\ }\href {https://link.aps.org/doi/10.1103/PhysRevLett.133.060602} {\bibfield  {journal} {\bibinfo  {journal} {Phys. Rev. Lett.}\ }\textbf {\bibinfo {volume} {133}},\ \bibinfo {pages} {060602} (\bibinfo {year} {2024})}\BibitemShut {NoStop}%
\bibitem [{\citenamefont {Larson}\ \emph {et~al.}(2025)\citenamefont {Larson}, \citenamefont {Yelton}, \citenamefont {Dodge}, \citenamefont {Okubo}, \citenamefont {Batarekh}, \citenamefont {Iaia}, \citenamefont {Kurinsky},\ and\ \citenamefont {Plourde}}]{Larson_PRXQuantum2025}%
  \BibitemOpen
  \bibfield  {author} {\bibinfo {author} {\bibfnamefont {C.}~\bibnamefont {Larson}}, \bibinfo {author} {\bibfnamefont {E.}~\bibnamefont {Yelton}}, \bibinfo {author} {\bibfnamefont {K.}~\bibnamefont {Dodge}}, \bibinfo {author} {\bibfnamefont {K.}~\bibnamefont {Okubo}}, \bibinfo {author} {\bibfnamefont {J.}~\bibnamefont {Batarekh}}, \bibinfo {author} {\bibfnamefont {V.}~\bibnamefont {Iaia}}, \bibinfo {author} {\bibfnamefont {N.}~\bibnamefont {Kurinsky}},\ and\ \bibinfo {author} {\bibfnamefont {B.}~\bibnamefont {Plourde}},\ }\href {https://link.aps.org/doi/10.1103/2lyd-8swv} {\bibfield  {journal} {\bibinfo  {journal} {PRX Quantum}\ }\textbf {\bibinfo {volume} {6}},\ \bibinfo {pages} {030339} (\bibinfo {year} {2025})}\BibitemShut {NoStop}%
\bibitem [{\citenamefont {Pan}\ \emph {et~al.}(2022)\citenamefont {Pan}, \citenamefont {Zhou}, \citenamefont {Yuan}, \citenamefont {Nie}, \citenamefont {Wei}, \citenamefont {Zhang}, \citenamefont {Li}, \citenamefont {Liu}, \citenamefont {Jiang}, \citenamefont {Catelani}, \citenamefont {Hu}, \citenamefont {Yan},\ and\ \citenamefont {Yu}}]{Pan_NatComm2022}%
  \BibitemOpen
  \bibfield  {author} {\bibinfo {author} {\bibfnamefont {X.}~\bibnamefont {Pan}}, \bibinfo {author} {\bibfnamefont {Y.}~\bibnamefont {Zhou}}, \bibinfo {author} {\bibfnamefont {H.}~\bibnamefont {Yuan}}, \bibinfo {author} {\bibfnamefont {L.}~\bibnamefont {Nie}}, \bibinfo {author} {\bibfnamefont {W.}~\bibnamefont {Wei}}, \bibinfo {author} {\bibfnamefont {L.}~\bibnamefont {Zhang}}, \bibinfo {author} {\bibfnamefont {J.}~\bibnamefont {Li}}, \bibinfo {author} {\bibfnamefont {S.}~\bibnamefont {Liu}}, \bibinfo {author} {\bibfnamefont {Z.~H.}\ \bibnamefont {Jiang}}, \bibinfo {author} {\bibfnamefont {G.}~\bibnamefont {Catelani}}, \bibinfo {author} {\bibfnamefont {L.}~\bibnamefont {Hu}}, \bibinfo {author} {\bibfnamefont {F.}~\bibnamefont {Yan}},\ and\ \bibinfo {author} {\bibfnamefont {D.}~\bibnamefont {Yu}},\ }\href {http://dx.doi.org/10.1038/s41467-022-34727-2} {\bibfield  {journal} {\bibinfo  {journal} {Nature Communications}\ }\textbf {\bibinfo {volume} {13}} (\bibinfo {year} {2022})}\BibitemShut {NoStop}%
\bibitem [{\citenamefont {Martinis}\ \emph {et~al.}(2009)\citenamefont {Martinis}, \citenamefont {Ansmann},\ and\ \citenamefont {Aumentado}}]{Martinis_PRL2009}%
  \BibitemOpen
  \bibfield  {author} {\bibinfo {author} {\bibfnamefont {J.~M.}\ \bibnamefont {Martinis}}, \bibinfo {author} {\bibfnamefont {M.}~\bibnamefont {Ansmann}},\ and\ \bibinfo {author} {\bibfnamefont {J.}~\bibnamefont {Aumentado}},\ }\href {https://link.aps.org/doi/10.1103/PhysRevLett.103.097002} {\bibfield  {journal} {\bibinfo  {journal} {Phys. Rev. Lett.}\ }\textbf {\bibinfo {volume} {103}},\ \bibinfo {pages} {097002} (\bibinfo {year} {2009})}\BibitemShut {NoStop}%
\bibitem [{\citenamefont {Uilhoorn}\ \emph {et~al.}(2021)\citenamefont {Uilhoorn}, \citenamefont {Kroll}, \citenamefont {Bargerbos}, \citenamefont {Nabi}, \citenamefont {Yang}, \citenamefont {Krogstrup}, \citenamefont {Kouwenhoven}, \citenamefont {Kou},\ and\ \citenamefont {de~Lange}}]{uilhoorn_arxiv2021}%
  \BibitemOpen
  \bibfield  {author} {\bibinfo {author} {\bibfnamefont {W.}~\bibnamefont {Uilhoorn}}, \bibinfo {author} {\bibfnamefont {J.~G.}\ \bibnamefont {Kroll}}, \bibinfo {author} {\bibfnamefont {A.}~\bibnamefont {Bargerbos}}, \bibinfo {author} {\bibfnamefont {S.~D.}\ \bibnamefont {Nabi}}, \bibinfo {author} {\bibfnamefont {C.-K.}\ \bibnamefont {Yang}}, \bibinfo {author} {\bibfnamefont {P.}~\bibnamefont {Krogstrup}}, \bibinfo {author} {\bibfnamefont {L.~P.}\ \bibnamefont {Kouwenhoven}}, \bibinfo {author} {\bibfnamefont {A.}~\bibnamefont {Kou}},\ and\ \bibinfo {author} {\bibfnamefont {G.}~\bibnamefont {de~Lange}},\ }\href {https://arxiv.org/abs/2105.11038} {\bibfield  {journal} {\bibinfo  {journal} {arXiv:2105.11038}\ } (\bibinfo {year} {2021})}\BibitemShut {NoStop}%
\bibitem [{\citenamefont {Erlandsson}\ \emph {et~al.}(2023)\citenamefont {Erlandsson}, \citenamefont {Sabonis}, \citenamefont {Kringh\o{}j}, \citenamefont {Larsen}, \citenamefont {Krogstrup}, \citenamefont {Petersson},\ and\ \citenamefont {Marcus}}]{Erlandsson_PRB2023}%
  \BibitemOpen
  \bibfield  {author} {\bibinfo {author} {\bibfnamefont {O.}~\bibnamefont {Erlandsson}}, \bibinfo {author} {\bibfnamefont {D.}~\bibnamefont {Sabonis}}, \bibinfo {author} {\bibfnamefont {A.}~\bibnamefont {Kringh\o{}j}}, \bibinfo {author} {\bibfnamefont {T.~W.}\ \bibnamefont {Larsen}}, \bibinfo {author} {\bibfnamefont {P.}~\bibnamefont {Krogstrup}}, \bibinfo {author} {\bibfnamefont {K.~D.}\ \bibnamefont {Petersson}},\ and\ \bibinfo {author} {\bibfnamefont {C.~M.}\ \bibnamefont {Marcus}},\ }\href {https://link.aps.org/doi/10.1103/PhysRevB.108.L121406} {\bibfield  {journal} {\bibinfo  {journal} {Phys. Rev. B}\ }\textbf {\bibinfo {volume} {108}},\ \bibinfo {pages} {L121406} (\bibinfo {year} {2023})}\BibitemShut {NoStop}%
\bibitem [{\citenamefont {Beenakker}(1992)}]{Beenakker_PRB1992}%
  \BibitemOpen
  \bibfield  {author} {\bibinfo {author} {\bibfnamefont {C.~W.~J.}\ \bibnamefont {Beenakker}},\ }\href {https://link.aps.org/doi/10.1103/PhysRevB.46.12841} {\bibfield  {journal} {\bibinfo  {journal} {Phys. Rev. B}\ }\textbf {\bibinfo {volume} {46}},\ \bibinfo {pages} {12841} (\bibinfo {year} {1992})}\BibitemShut {NoStop}%
\bibitem [{\citenamefont {Aguado}(2020)}]{Aguado_APL2020}%
  \BibitemOpen
  \bibfield  {author} {\bibinfo {author} {\bibfnamefont {R.}~\bibnamefont {Aguado}},\ }\href {http://dx.doi.org/10.1063/5.0024124} {\bibfield  {journal} {\bibinfo  {journal} {Applied Physics Letters}\ }\textbf {\bibinfo {volume} {117}} (\bibinfo {year} {2020})}\BibitemShut {NoStop}%
\bibitem [{\citenamefont {Vaitiekėnas}\ \emph {et~al.}(2020)\citenamefont {Vaitiekėnas}, \citenamefont {Winkler}, \citenamefont {van Heck}, \citenamefont {Karzig}, \citenamefont {Deng}, \citenamefont {Flensberg}, \citenamefont {Glazman}, \citenamefont {Nayak}, \citenamefont {Krogstrup}, \citenamefont {Lutchyn},\ and\ \citenamefont {Marcus}}]{Vaitiek_Science2020}%
  \BibitemOpen
  \bibfield  {author} {\bibinfo {author} {\bibfnamefont {S.}~\bibnamefont {Vaitiekėnas}}, \bibinfo {author} {\bibfnamefont {G.~W.}\ \bibnamefont {Winkler}}, \bibinfo {author} {\bibfnamefont {B.}~\bibnamefont {van Heck}}, \bibinfo {author} {\bibfnamefont {T.}~\bibnamefont {Karzig}}, \bibinfo {author} {\bibfnamefont {M.-T.}\ \bibnamefont {Deng}}, \bibinfo {author} {\bibfnamefont {K.}~\bibnamefont {Flensberg}}, \bibinfo {author} {\bibfnamefont {L.~I.}\ \bibnamefont {Glazman}}, \bibinfo {author} {\bibfnamefont {C.}~\bibnamefont {Nayak}}, \bibinfo {author} {\bibfnamefont {P.}~\bibnamefont {Krogstrup}}, \bibinfo {author} {\bibfnamefont {R.~M.}\ \bibnamefont {Lutchyn}},\ and\ \bibinfo {author} {\bibfnamefont {C.~M.}\ \bibnamefont {Marcus}},\ }\href {http://dx.doi.org/10.1126/science.aav3392} {\bibfield  {journal} {\bibinfo  {journal} {Science}\ }\textbf {\bibinfo {volume} {367}} (\bibinfo {year} {2020})}\BibitemShut {NoStop}%
\bibitem [{\citenamefont {Glazman}\ and\ \citenamefont {Catelani}(2021)}]{Glazman_SciPost2021}%
  \BibitemOpen
  \bibfield  {author} {\bibinfo {author} {\bibfnamefont {L.~I.}\ \bibnamefont {Glazman}}\ and\ \bibinfo {author} {\bibfnamefont {G.}~\bibnamefont {Catelani}},\ }\href {https://scipost.org/10.21468/SciPostPhysLectNotes.31} {\bibfield  {journal} {\bibinfo  {journal} {SciPost Phys. Lect. Notes}\ ,\ \bibinfo {pages} {31}} (\bibinfo {year} {2021})}\BibitemShut {NoStop}%
\bibitem [{\citenamefont {Vaitiek\ifmmode~\dot{e}\else \.{e}\fi{}nas}\ \emph {et~al.}(2018)\citenamefont {Vaitiek\ifmmode~\dot{e}\else \.{e}\fi{}nas}, \citenamefont {Deng}, \citenamefont {Nyg\aa{}rd}, \citenamefont {Krogstrup},\ and\ \citenamefont {Marcus}}]{SolePRL_2018}%
  \BibitemOpen
  \bibfield  {author} {\bibinfo {author} {\bibfnamefont {S.}~\bibnamefont {Vaitiek\ifmmode~\dot{e}\else \.{e}\fi{}nas}}, \bibinfo {author} {\bibfnamefont {M.-T.}\ \bibnamefont {Deng}}, \bibinfo {author} {\bibfnamefont {J.}~\bibnamefont {Nyg\aa{}rd}}, \bibinfo {author} {\bibfnamefont {P.}~\bibnamefont {Krogstrup}},\ and\ \bibinfo {author} {\bibfnamefont {C.~M.}\ \bibnamefont {Marcus}},\ }\href {https://doi.org/10.1103/PhysRevLett.121.037703} {\bibfield  {journal} {\bibinfo  {journal} {Phys. Rev. Lett.}\ }\textbf {\bibinfo {volume} {121}},\ \bibinfo {pages} {037703} (\bibinfo {year} {2018})}\BibitemShut {NoStop}%
\bibitem [{\citenamefont {Bargerbos}\ \emph {et~al.}(2023)\citenamefont {Bargerbos}, \citenamefont {Splitthoff}, \citenamefont {Pita-Vidal}, \citenamefont {Wesdorp}, \citenamefont {Liu}, \citenamefont {Krogstrup}, \citenamefont {Kouwenhoven}, \citenamefont {Andersen},\ and\ \citenamefont {Grünhaupt}}]{Bargerbos_PRApplied2023}%
  \BibitemOpen
  \bibfield  {author} {\bibinfo {author} {\bibfnamefont {A.}~\bibnamefont {Bargerbos}}, \bibinfo {author} {\bibfnamefont {L.~J.}\ \bibnamefont {Splitthoff}}, \bibinfo {author} {\bibfnamefont {M.}~\bibnamefont {Pita-Vidal}}, \bibinfo {author} {\bibfnamefont {J.~J.}\ \bibnamefont {Wesdorp}}, \bibinfo {author} {\bibfnamefont {Y.}~\bibnamefont {Liu}}, \bibinfo {author} {\bibfnamefont {P.}~\bibnamefont {Krogstrup}}, \bibinfo {author} {\bibfnamefont {L.~P.}\ \bibnamefont {Kouwenhoven}}, \bibinfo {author} {\bibfnamefont {C.~K.}\ \bibnamefont {Andersen}},\ and\ \bibinfo {author} {\bibfnamefont {L.}~\bibnamefont {Grünhaupt}},\ }\href {http://dx.doi.org/10.1103/PhysRevApplied.19.024014} {\bibfield  {journal} {\bibinfo  {journal} {Physical Review Applied}\ }\textbf {\bibinfo {volume} {19}} (\bibinfo {year} {2023})}\BibitemShut {NoStop}%
\bibitem [{\citenamefont {McEwen}\ \emph {et~al.}(2024)\citenamefont {McEwen} \emph {et~al.}}]{McEwen_PRL2024}%
  \BibitemOpen
  \bibfield  {author} {\bibinfo {author} {\bibfnamefont {M.}~\bibnamefont {McEwen}} \emph {et~al.},\ }\href {https://doi.org/10.1103/PhysRevLett.133.240601} {\bibfield  {journal} {\bibinfo  {journal} {Physical Review Letters}\ }\textbf {\bibinfo {volume} {133}},\ \bibinfo {pages} {240601} (\bibinfo {year} {2024})}\BibitemShut {NoStop}%
\end{thebibliography}%

\end{document}